\documentclass[aps,prd,twocolumn,superscriptaddress,nofootinbib,floatfix]{revtex4-2}

\usepackage[utf8]{inputenc}
\usepackage[T1]{fontenc}
\usepackage{lmodern}
\usepackage{amsmath,amssymb,mathtools,bm}
\usepackage{graphicx}
\usepackage{booktabs}
\usepackage{multirow}
\usepackage{xcolor}
\usepackage{url}
\usepackage{microtype}
\usepackage{tikz}
\usetikzlibrary{arrows.meta,positioning,fit,backgrounds,calc}
\usepackage[colorlinks=true,linkcolor=blue,citecolor=blue,urlcolor=blue]{hyperref}
\usepackage{cleveref}

\definecolor{cTh}{HTML}{3D5A80}
\definecolor{cCl}{HTML}{2A9D8F}
\definecolor{cIn}{HTML}{BC6C25}
\definecolor{cDa}{HTML}{5C6670}

\crefname{equation}{Equation}{Equations}
\Crefname{equation}{Equation}{Equations}
\crefname{figure}{Figure}{Figures}
\Crefname{figure}{Figure}{Figures}
\crefname{table}{Table}{Tables}
\Crefname{table}{Table}{Tables}
\crefname{section}{Section}{Sections}
\Crefname{section}{Section}{Sections}
\crefname{subsection}{Subsection}{Subsections}
\Crefname{subsection}{Subsection}{Subsections}
\crefname{subsubsection}{Subsection}{Subsections}
\Crefname{subsubsection}{Subsection}{Subsections}
\crefname{appendix}{Appendix}{Appendices}
\Crefname{appendix}{Appendix}{Appendices}

\newcommand{\mpl}{M_{\mathrm{Pl}}}
\newcommand{\Trh}{T_{\mathrm{rh}}}
\newcommand{\Trhmax}{T_{\mathrm{rh}}^{\max}}
\newcommand{\TBBN}{T_{\mathrm{BBN}}}
\newcommand{\wrh}{w_{\mathrm{rh}}}
\newcommand{\Nrh}{N_{\mathrm{rh}}}
\newcommand{\Nstar}{N_*}
\newcommand{\grh}{g_{\mathrm{rh}}}
\newcommand{\gsrh}{g_{s,\mathrm{rh}}}
\newcommand{\ns}{n_s}

\begin{document}

\title{\boldmath Reheating Constraints on the $\alpha$-Starobinsky Model \\ and Implications for Inflation Decays}

\author{Karim Carrion}
\email{kcarrion@fisica.unam.mx}
\affiliation{Instituto de F\'isica, Universidad Nacional Aut\'onoma de M\'exico, Circuito de la Investigaci\'on Cient\'ifica, Ciudad Universitaria, Cd.\ de M\'exico, 04510, M\'exico}
\affiliation{Instituto de Ciencias F\'isicas, Universidad Nacional Aut\'onoma de M\'exico, Av.\ Universidad s/n, Cuernavaca, Morelos, 62210, M\'exico}

\author{Luis E. Padilla}
\email{lepadilla@rikkyo.ac.jp}
\affiliation{Department of Physics, Rikkyo University, Tokyo 171-8501, Japan}

\author{Gabriel Germ\'an}
\affiliation{Instituto de Ciencias F\'isicas, Universidad Nacional Aut\'onoma de M\'exico, Av.\ Universidad s/n, Cuernavaca, Morelos, 62210, M\'exico}

\author{Juan Carlos Hidalgo}
\affiliation{Instituto de Ciencias F\'isicas, Universidad Nacional Aut\'onoma de M\'exico, Av.\ Universidad s/n, Cuernavaca, Morelos, 62210, M\'exico}


\date{\today}

\begin{abstract}
We constrain the $\alpha$-Starobinsky model of inflation with an explicit reheating sector, in which the number of $e$-folds $\Nstar$ is computed for every sampled point from the reheating equation of state $\wrh$ and temperature $\Trh$, using a fully numerical pivot-scale matching in \texttt{CLASS}. We analyze six combinations of Planck, Atacama Cosmology Telescope (ACT) DR6 and DESI DR2 baryon acoustic oscillation (BAO) data. The CMB sees reheating only through $\ln R_{\rm rad}\simeq\Nstar-N_{\rm inst}$, the shift of $\Nstar$ from its instantaneous-reheating value. At $\alpha=1$, Planck gives $\ns=0.9656^{+0.0015}_{-0.0017}$ and is consistent with instantaneous reheating, $\ln R_{\rm rad}=0.4\pm2.7$. Planck, ACT and DESI together (P-ACT+DESI) give $\ns=0.9689^{+0.0012}_{-0.0016}$ and $\ln R_{\rm rad}=6.5^{+2.5}_{-3.1}$, which prefers a stiff phase ($\wrh>0.47$ at 95\%). With $\alpha$ free they give $\log_{10}\alpha=1.0^{+0.5}_{-0.8}$, $\ns=0.9706^{+0.0016}_{-0.0020}$ and $\ln R_{\rm rad}=6^{+3}_{-4}$. This preference is not in agreement with the bound on the primordial gravitational-wave contribution. Imposing that bound approximately removes about half of the preferred stiff region. Comparing to an otherwise identical analysis in which $\Nstar$ is a free parameter, we find that the reheating sector acts mainly as a prior on $\Nstar$: at $\alpha=1$ it narrows the inflationary posteriors, and with $\alpha$ free it removes the largest values of $\Nstar$ and doubles the preferred $r$ for P-ACT+DESI. Finally, we briefly discuss the connection between these phenomenological constraints and representative perturbative inflaton decay scenarios, highlighting the limitations of this interpretation for stiff reheating histories.
\end{abstract}

\maketitle

\section{Introduction}
Inflation was proposed to explain why the observable Universe is so homogeneous and spatially flat~\cite{Guth1981,Linde1982,AlbrechtSteinhardt1982}. It also provides a dynamical origin for the primordial fluctuations: quantum modes of the inflaton are stretched beyond the horizon, freeze out as curvature perturbations, and later appear as Cosmic Microwave Background (CMB) anisotropies and as the seeds of large-scale structure~\cite{MukhanovChibisov1981,Hawking1982,Starobinsky1982,GuthPi1982,BardeenSteinhardtTurner1983}. These ideas have been extensively reviewed and confronted with data~\cite{EncyclopaediaInflationaris,MartinRingevalVennin2024}.

Of the many inflationary scenarios that have survived increasingly precise CMB constraints, the Starobinsky model~\cite{Starobinsky1980} remains a useful reference point. Originally formulated as an $R^2$ extension of General Relativity, it can be recast in the Einstein frame as a canonical scalar field rolling down an asymptotically flat potential. In this frame, its predictions for the scalar spectral index $\ns$ and the tensor-to-scalar ratio $r$ lie within the observationally favored region~\cite{PlanckInflation2020}. The model belongs to the broader class of $\alpha$-attractors~\cite{KalloshLinde2013,KalloshLindeRoest2013,KalloshLindeRoest2014}, whose universal behavior makes them predictive and which arise naturally in supergravity~\cite{FKLP2013}. Thus, the $\alpha$-Starobinsky generalization is a natural benchmark against which to test the impact of early universe observations.

Recent primary-CMB measurements from the sixth data release (DR6) of the Atacama Cosmology Telescope (ACT)~\cite{Louis2025,Calabrese2025} and from the South Pole Telescope (SPT-3G)~\cite{Camphuis2025}, together with baryon acoustic oscillation (BAO) measurements from the second data release (DR2) of the Dark Energy Spectroscopic Instrument (DESI)~\cite{DESIDR2}, have renewed interest in testing plateau models of inflation. Planck and ACT probe complementary multipole ranges of the primary CMB, while BAO data break degeneracies in the late-time sector. Although BAO do not constrain $\ns$ directly, they impact the late-time parameters with which $\ns$ is correlated in a CMB fit, and therefore the inferred value of $\ns$. The combination of Planck, ACT DR6, CMB lensing and DESI DR1 BAO (P-ACT-LB) gives $\ns=0.9743\pm0.0034$~\cite{Louis2025}, which places the canonical Starobinsky model ($\alpha=1$) and other attractor potentials in $\simeq2\sigma$ tension with this combination~\cite{KalloshLindeRoest2025,DreesXu2025,StarobinskyTension2026}. This has motivated deformations and extensions of the Starobinsky potential~\cite{ACTConstraints2025,QuantumCorrected2025,MinimalExtensions2026}, as well as a consideration of the post-inflationary history~\cite{Zharov2025,Haque2025,Mondal2026,Ellis2025}. For plateau models a larger $\ns$ corresponds to a larger number of $e$-folds $\Nstar$ between the horizon exit of the pivot scale and the end of inflation, so the question is directly tied to how long reheating lasts and how the Universe evolves in this period.

The value of $\Nstar$ is not an independent quantity: it is linked to the expansion history from the end of inflation to the onset of radiation domination~\cite{LiddleLeach2003,DodelsonHui2003,MartinRingeval2010}. Different reheating histories displace the same observed scale to different locations on the potential, shifting the associated values of $\ns$ and $r$~\cite{DaiKamionkowskiWang2014,Cook2015,Lozanov2019}. Usually the duration of reheating is controlled by the effective equation of state $\wrh$ of the dominant component during reheating and by the reheating temperature $\Trh$ at which radiation domination begins. Yet other parameters may come into play depending on the underlying reheating model.

The impact of reheating uncertainties on inflationary observables has been explored in broad surveys of models~\cite{Ballardini2025}, for instance, in $\alpha$-attractor and $\alpha$-Starobinsky realizations~\cite{UenoYamamoto2016,ReheatingDynamics2025,Iacconi2025}, in D-brane models~\cite{Yi2006}, in radiative plateau inflation~\cite{BallesterosTamarit2016}, and in connection with dark-matter production during reheating~\cite{Garcia2020}. These studies show that the post-inflationary history can substantially broaden the region accessible in the $(\ns,r)$ plane. In particular, some T-models can sustain an extended reheating stage with a stiff equation of state ($\wrh>1/3$), which raises $\ns$ toward the ACT and DESI preferred values~\cite{Iacconi2025}, whereas such a stage enhances the primordial gravitational-wave background and is constrained by the bound on extra radiation~(e.g.~\cite{Dorsch2026,Eroncel2025}). Reheating can also proceed through other channels, such as the evaporation of primordial black holes~\cite{Hidalgo:2011fj}.

Our previous analysis of the $\alpha$-Starobinsky model~\cite{Carrion2026}, based on Planck 2018 temperature and polarization data, ACT DR6 CMB lensing\footnote{CMB lensing was found to have no impact on the inflationary constraints.}~\cite{ACTDR6Qu,ACTDR6Madhavacheril} and DESI DR2 BAO, treated $\Nstar$ as a free parameter. Adding DESI to Planck shifted the posterior from $\Nstar=56.7^{+5.2}_{-8.3}$ to $63.4^{+5.5}_{-8.2}$ at $\alpha=1$ and, once $\alpha$ was free, toward $\log_{10}\alpha>0$. A free $\Nstar$, however, is a phenomenological stand-in for the reheating epoch rather than the prediction of a specific post-inflationary history, and the associated constraints need not survive once reheating is modeled explicitly.

In the present paper we promote $\Nstar$ from a free parameter to a derived quantity through an explicit reheating sector. The post-inflationary stage is characterized by an effective equation of state $\wrh$ and a reheating temperature $\Trh$, which directly determine $\Nstar$. For every sampled point, the inflationary background computed by \texttt{CLASS} determines $\phi_*$ and $\Nstar$ before the scalar and tensor perturbations are evolved, so that, unlike in our previous analysis, no slow-roll consistency relation is imposed in the inference process.

Our analysis consists of four main steps. First, we constrain the canonical and generalized models with six combinations of Planck, ACT and DESI data, and identify the single combination of reheating parameters preferred by the data: the departure $\ln R_{\rm rad}\simeq\Nstar-N_{\rm inst}$ of $\Nstar$ from its value for instantaneous reheating, $N_{\rm inst}$. Second, since the CMB likelihood depends on $(\wrh,\Trh)$ only through $\Nstar$, we compare with our previous approach, in which $\Nstar$ is a free parameter, and separate the effect of the reheating prior from the information supplied by the data. Third, we restrict the equation of state to the non-stiff range $\wrh\le1/3$ and show how the high-$\ns$ combinations respond when a stiff post-inflationary phase is forbidden, including the constraint on such a phase from the primordial gravitational-wave background. Finally, we illustrate how the inferred reheating temperatures map onto representative perturbative inflaton decay rates, while distinguishing this conditional interpretation from a microscopic realization of stiff reheating.

The paper is organized as follows. \Cref{sec:framework} reviews the inflationary dynamics and introduces the reheating matching. \Cref{sec:method} describes the numerical implementation, the data and the priors. The results for the canonical and generalized models are presented in \cref{sec:results}, together with two tests: a comparison with an analysis without a reheating sector and a restrictive reheating prior. \Cref{sec:microphysics} presents illustrative perturbative decay scenarios, and \cref{sec:conclusions} summarizes our conclusions.

\section{Inflationary and reheating framework}
\label{sec:framework}

\subsection{Inflationary background}

The $\alpha$-Starobinsky model generalizes the Starobinsky potential by introducing a parameter $\alpha>0$ that controls the width of the plateau. It belongs to the E-model class of $\alpha$-attractors, with potential
\begin{equation}
\begin{gathered}
V(\phi)=V_0 f(\phi),\\ f(\phi)=\left(1-e^{-b\phi/\mpl}\right)^2, \qquad
b\equiv\sqrt{\frac{2}{3\alpha}},
\end{gathered}
\label{eq:potential}
\end{equation}
where $V_0$ sets the amplitude of the scalar spectrum and $\mpl=(8\pi G)^{-1/2}$ is the reduced Planck mass. The canonical Starobinsky model is recovered for $\alpha=1$.

In our numerical analysis, the number of $e$-folds between horizon exit and the end of inflation is obtained from the exact background,
\begin{equation}
\Nstar\equiv\ln\frac{a_{\rm end}}{a_*}=\int_{t_*}^{t_{\rm end}}H\,dt,
\label{eq:Nbg}
\end{equation}
with the end of inflation defined by $\ddot a=0$, i.e.\ $\epsilon_H\equiv-\dot H/H^2=1$. Scalar and tensor perturbations are then evolved mode by mode on this background.

For comparison with semi-analytical treatments, the potential slow-roll parameters $\epsilon_V = \tfrac{1}{2}\mpl^2(V'/V)^2$ and $\eta_V=\mpl^2V''/V$ give $\ns\simeq1-6\epsilon_V+2\eta_V$ and $r\simeq16\epsilon_V$. In this approximation inflation ends at
\begin{equation}
\phi_{\rm end}^{\rm SR}=\sqrt{\frac{3\alpha}{2}}\,\mpl\ln\left(1+\frac{2}{\sqrt{3\alpha}}\right),
\label{eq:phiend}
\end{equation}
and the number of $e$-folds for a pivot field value $\phi_*$ is
\begin{equation}
N_*^{\rm SR}=\frac{1}{\mpl^2}\int_{\phi_{\rm end}^{\rm SR}}^{\phi_*}\frac{V}{V'}\,d\phi.
\label{eq:NSR}
\end{equation}
For $\Nstar\gg1$ and $\alpha\lesssim\Nstar$ they reduce to the relations shared by all $\alpha$-attractors, of which the $\alpha$-Starobinsky (E-model) potential is one example,
\begin{equation}
\ns\simeq1-\frac{2}{\Nstar},\qquad
r\simeq\frac{12\alpha}{\Nstar^2},\qquad
A_s\simeq\frac{V_0\Nstar^2}{18\pi^2\alpha\,\mpl^4}.
\label{eq:attractor}
\end{equation}
These expressions are used for the analytical benchmark described below and for estimates, but not in the inference. For $\alpha=1$, slow roll places the end of inflation at $\phi_{\rm end}^{\rm SR}\simeq0.94\,\mpl$, whereas $\ddot a=0$ occurs later, at $\phi_{\rm end}\simeq0.61\,\mpl$. Every quantity entering the numerical matching, described below, is evaluated at the latter endpoint.

\subsection{Effective description of reheating}
\label{subsec:effective-reheating}

The post-inflationary reheating is described by the reheating temperature $\Trh$, defined as the temperature of the thermal bath at the onset of radiation domination, and an effective equation-of-state parameter $\wrh$. For a canonical scalar oscillating around a minimum $V\propto\phi^n$, the single-oscillation average gives $\langle w\rangle=(n-2)/(n+2)$~\cite{Turner1983}: $\langle w\rangle=0$ for a quadratic minimum and $1/3$ for a quartic one. Near its minimum the $\alpha$-Starobinsky potential is quadratic, so matter-like reheating ($\wrh=0$) is the canonical reference case.

This value need not describe the entire post-inflationary epoch: particle production, fragmentation, nonperturbative resonance, backreaction, rescattering and the gradual build-up of relativistic species can modify the background evolution before thermalization is complete~\cite{Lozanov2019}. In perturbative decay scenarios, radiation produced during reheating also gives a small, generally nonzero contribution to the $e$-fold-averaged equation of state of the combined system~\cite{Ellis2015}. For a general time-dependent equation of state $w(N)$, energy conservation gives $\ln(\rho_{\rm end}/\rho_{\rm rh})=3\int_{\rm reh}[1+w(N)]\,dN$. Since $\Nrh=\int_{\rm reh}dN$, the matching depends on $w(N)$ through its $e$-fold average,
\begin{equation}
\wrh^{\rm eff}\equiv\frac{1}{\Nrh}\int_{\rm reh}w(N)\,dN.
\label{eq:weff}
\end{equation}
Throughout the analysis the constant $\wrh$ represents this effective average~\cite{MunozKamionkowski2015}, not necessarily the instantaneous equation of state. It should be distinguished from the single-oscillation average above. Other observables, such as the primordial gravitational-wave spectrum, can depend on the full history $w(N)$.

With this effective description the reheating duration is
\begin{equation}
\Nrh=\frac{1}{3(1+\wrh)}\ln\left(\frac{\rho_{\rm end}}{\rho_{\rm rh}}\right),
\label{eq:Nrh}
\end{equation}
where $\rho_{\rm end}$ is the energy density at the end of inflation and $\rho_{\rm rh}=(\pi^2/30)\,\grh\Trh^4$ is the energy density at the onset of radiation domination. Here $\grh$ ($\gsrh$) denotes the effective number of relativistic energy (entropy) degrees of freedom at reheating. The maximal temperature corresponds to instantaneous reheating, $\Nrh=0$, for which $\rho_{\rm rh}=\rho_{\rm end}$:
\begin{equation}
\Trhmax=\left(\frac{30\,\rho_{\rm end}}{\pi^2\grh}\right)^{1/4}.
\label{eq:Tmax}
\end{equation}
In our analysis, we set $\grh=\gsrh=106.75$. At low reheating temperatures the actual values are smaller, decreasing to $\simeq10.75$ near Big Bang nucleosynthesis (BBN)~\cite{Saikawa2018,Husdal2016}. Since $\Nstar$ depends on $\grh=\gsrh$ through $-[\wrh/3(1+\wrh)]\ln\grh$, using $10.75$ instead of $106.75$ for a stiff phase with $\wrh=0.75$ would increase $\Nstar$ by $\simeq0.3$ and $\ns$ by $\simeq2\times10^{-4}$, about $0.1\sigma$ of our tightest constraint.

\subsection{Pivot-scale matching and numerical validation}
\label{subsec:reheating}

Following the pivot scale, $k_*=a_*H_*$, from horizon exit through reheating and into the radiation-dominated era gives~\cite{LiddleLeach2003,DaiKamionkowskiWang2014,Cook2015,MunozKamionkowski2015}
\begin{equation}
\begin{split}
\frac{k_*}{a_0H_*}&=\frac{a_*}{a_{\rm end}}\,\frac{a_{\rm end}}{a_{\rm rh}}\,\frac{a_{\rm rh}}{a_0}\\
&=e^{-\Nstar-\Nrh}\left(\frac{g_{s,0}}{\gsrh}\right)^{1/3}\frac{T_0}{\Trh},
\end{split}
\label{eq:epochs}
\end{equation}
where $\Nrh$ is given by \cref{eq:Nrh} and the last factor follows from entropy conservation after reheating, $g_s a^3T^3={\rm const}$. Here $g_{s,0}=43/11$ is the present number of entropy degrees of freedom, and $a_0$ and $T_0$ are the present scale factor and CMB temperature. Taking the logarithm gives
\begin{equation}
\begin{split}
N_*^{\rm match}={}&\ln\frac{H_*}{k_*}+\frac13\ln\frac{g_{s,0}}{\gsrh}+\ln\frac{a_0T_0}{\Trh}\\
&-\frac{1}{3(1+\wrh)}\ln\frac{\rho_{\rm end}}{\rho_{\rm rh}},
\end{split}
\label{eq:matching-numerical}
\end{equation}
with $H_*$ and $\rho_{\rm end}=3H_{\rm end}^2\mpl^2$ evaluated on the numerical background.

For potentials of the form $V(\phi)=V_0 f(\phi)$, eliminating $V_0$ with the slow-roll normalization $A_s\simeq V_*/(24\pi^2\mpl^4\epsilon_V)$ and approximating $\rho_{\rm end}\simeq\frac32V(\phi_{\rm end})$ turns \cref{eq:matching-numerical} into the semi-analytical relation~\cite{ReheatingDynamics2025,GermanHidalgo2025}
\begin{align}
N_*^{\rm match}
={}&
\ln\left[\frac{2}{k_*}\left(\frac{g_{s,0}}{\gsrh}\right)^{1/3}\pi\sqrt{A_s}\,a_0T_0\right]
\nonumber\\
&+\frac{1}{3(1+\wrh)}\ln\left(\frac{\grh}{540A_s}\right)
\nonumber\\
&+\frac{1-3\wrh}{3(1+\wrh)}\ln\left(\frac{\Trh}{\mpl}\right)
\nonumber\\
&+\frac{1}{3(1+\wrh)}\ln\left[\frac{\left(\mpl f'(\phi_*)\right)^{1+3\wrh}}{f(\phi_*)^{3\wrh}f(\phi_{\rm end})}\right],
\label{eq:matching}
\end{align}
where primes denote derivatives with respect to $\phi$ and $k_*/(a_0T_0)$ is evaluated in consistent units.

In the inference, the pivot field value is fixed by the matching condition
\begin{equation}
N_*^{\rm bg}(\phi_*)=N_*^{\rm match}(\phi_*),
\label{eq:matching-condition}
\end{equation}
where $N_*^{\rm bg}$ is the exact number of $e$-folds of \cref{eq:Nbg} and $N_*^{\rm match}$ is given by \cref{eq:matching-numerical}. We solve \cref{eq:matching-condition} with a bracketed root search, each step requiring one background integration. The residual is limited by the background step, $\lesssim5\times10^{-3}$ $e$-folds, corresponding to $|\Delta\ns|\lesssim10^{-5}$, so no slow-roll approximation enters the likelihood. Varying $\Trh$ or $\wrh$ displaces $\phi_*$ along the inflationary trajectory and thereby changes $\Nstar$ and the spectra evaluated at horizon exit.

As an independent check of the matching implementation, we reproduced the semi-analytical $\alpha$-Starobinsky reheating calculation of~\cite{ReheatingDynamics2025} using \cref{eq:matching}, and compared it with the numerical implementation of \cref{eq:matching-numerical}. Over the physical range of $(\wrh,\Trh)$ and $0.1\le\alpha\le10$, the two matched values of $\Nstar$ agree to within $0.3$ $e$-folds. At fixed $\phi_*$, however, the slow-roll count of \cref{eq:NSR} falls short of the exact one, \cref{eq:Nbg}, by about one $e$-fold, so the semi-analytical $\phi_*$ lies about one $e$-fold higher on the potential and its slow-roll value of $\ns$ is larger by $(5$--$10)\times10^{-4}$ for $50\lesssim\Nstar\lesssim70$ at $\alpha=1$, a third to two thirds of our tightest $\sigma(\ns)$. Only the full-background matching enters the inference.

The matching also identifies the combination of reheating parameters to which the CMB is sensitive. Defining the reheating parameter $R_{\rm rad}$~\cite{MartinRingeval2010}, we obtain
\begin{equation}
\begin{split}
\ln R_{\rm rad}&\equiv\frac{1-3\wrh}{12(1+\wrh)}\ln\frac{\rho_{\rm rh}}{\rho_{\rm end}}\\
&=\frac{1-3\wrh}{3(1+\wrh)}\ln\frac{\Trh}{\Trhmax}
\simeq\Nstar-N_{\rm inst},
\end{split}
\label{eq:DeltaN}
\end{equation}
where $N_{\rm inst}$ is the number of $e$-folds for instantaneous reheating, computed with the same numerical procedure by setting $\Trh=\Trhmax$. The expression for $\ln R_{\rm rad}$ in terms of reheating parameters is exact within a constant effective $\wrh$ description. Its identification with $\Nstar-N_{\rm inst}$ is approximate, because the matched solutions generally have slightly different $\phi_*$ and $H_*$. During the analysis, $\ln R_{\rm rad}$ is evaluated sample by sample from the temperature expression in \cref{eq:DeltaN}, using the numerical $\Trhmax$ of each sample.

The instantaneous solution depends on the potential shape and only logarithmically on its height, $\partial N_{\rm inst}/\partial\ln V_0\simeq1/4$ at fixed $\alpha$. For $A_s=2.1\times10^{-9}$ we find $N_{\rm inst}=55.7$ at $\alpha=1$, rising from $55.0$ to $56.5$ between $\alpha=0.1$ and $10$.

\Cref{eq:DeltaN} is the key relation of this work. The two reheating parameters affect the CMB through one number, $\ln R_{\rm rad}$: all pairs $(\wrh,\Trh)$ with the same $\ln R_{\rm rad}$ give the same $\Nstar$ and the same spectra. The data therefore select a curve of constant $\ln R_{\rm rad}$ in the $(\wrh,\Trh)$ plane, but not a point along it. The line $\ln R_{\rm rad}=0$ includes both instantaneous reheating ($\Trh=\Trhmax$) and radiation-like reheating ($\wrh=1/3$), where the matching is independent of $\Trh$, as seen in \Cref{eq:matching}. For $\wrh\le1/3$ we have $\ln R_{\rm rad}\le0$, so $\Nstar$ cannot exceed its instantaneous-reheating value $N_{\rm inst}$ and positive values require a stiff phase. The marginal constraints on $\Trh$, and to a lesser extent on $\wrh$, therefore depend on the prior induced by the uniform priors on $\wrh$ and $\log_{10}\Trh$ and by the physical limits on $\Trh$ discussed below.

\subsection{Physical range and impact of the reheating parameters}
\label{sec:reheating-priors-origin}

For the inference we adopt the permissive range $-1/3<\wrh\le1$, whose lower boundary excludes a further accelerated stage and whose upper value is the stiff limit. We also consider the restrictive range $-1/3<\wrh\le1/3$, which excludes stiff expansion. Unless stated otherwise, posteriors refer to the permissive interval, and the restrictive case is studied separately later on. This phenomenological treatment follows previous reheating analyses~\cite{GermanHidalgo2025,Ballardini2025}.

Values $\wrh>1/3$ require a sufficiently stiff post-inflationary expansion history. Although the $\alpha$-Starobinsky potential is quadratic near its minimum and leads to matter-like oscillations, a short kinetic-dominated transient can occur as the inflaton approaches the minimum. Such a transient does not, however, necessarily produce an effective reheating equation of state above $1/3$. Sustaining a stiff expansion history over a sufficiently long interval may require additional post-inflationary dynamics beyond the conventional oscillatory reheating scenario. The onset of radiation domination also depends on the mechanism responsible for radiation production, which is not specified by our effective parametrization. We therefore treat $\wrh>1/3$ phenomenologically, without assuming a particular post-inflationary evolution.

A stiff phase also affects the primordial gravitational-wave background, providing an additional constraint beyond the CMB fit. Inflation produces tensor modes with an almost flat spectrum. Once a mode re-enters the horizon, its energy density redshifts like radiation, $\rho_{\rm GW}\propto a^{-4}$. During a stiff phase the background redshifts faster, $\rho\propto a^{-3(1+\wrh)}$ with $\wrh>1/3$, enhancing the relative contribution of modes that re-enter during this period. For an approximately constant equation of state, this produces a blue spectrum, $\Omega_{\rm GW}(f)\propto f^{2(3\wrh-1)/(3\wrh+1)}$~\cite{Giovannini1998,FigueroaTanin2019}. The enhancement depends on the duration and stiffness of the phase, and therefore on the detailed post-inflationary expansion history. Because gravitational waves behave as radiation, this background contributes to the effective number of relativistic species, $N_{\rm eff}$, at BBN and recombination. In particular, $\Delta N_{\rm eff}\simeq\int\Omega_{\rm GW}h^2\,d\ln f/(5.6\times10^{-6})$~\cite{BoyleSteinhardt2008}. We do not include this bound in the likelihood. Instead, we impose $\Delta N_{\rm eff}<0.3$, roughly the $95\%$ upper bound from Planck~\cite{Planck2018VI}, on the permissive chains afterwards, estimating $\Omega_{\rm GW}$ sample by sample under the approximation of a constant effective equation of state and a transfer function of order unity. Since the effective reheating parametrization does not uniquely determine the gravitational-wave spectrum for a time-dependent equation of state, the resulting exclusion should be regarded as an order-of-magnitude estimate.

The reheating temperature is bounded from below by the requirement that radiation domination be established before BBN and from above by $\Trhmax$, \cref{eq:Tmax}. In the numerical analysis $\rho_{\rm end}=\frac32V$ is evaluated at $\ddot a=0$, where this relation is exact. The slow-roll estimate,
\begin{equation}
\rho_{\rm end}\simeq\frac32V(\phi_{\rm end}^{\rm SR})=\frac{6\,V_0}{\left(2+\sqrt{3\alpha}\right)^2},
\label{eq:rhoend}
\end{equation}
overestimates the numerical $\Trhmax$ by about $0.07$~dex\footnote{One dex is a factor of ten, i.e.\ a unit change in $\log_{10}$.} at $\alpha=1$. With the numerical endpoint we find $\log_{10}(\Trhmax/{\rm GeV})=15.350\pm0.010$ for $\alpha=1$. At fixed $A_s$, $V_0\propto\alpha$ to leading order, so $\Trhmax\propto[\alpha/(2+\sqrt{3\alpha})^2]^{1/4}$. With the numerical endpoint, $\log_{10}(\Trhmax/{\rm GeV})$ varies from $15.20$ to $15.42$ over $0.1\le\alpha\le10$. Configurations with $\Trh>\Trhmax$ would give $\Nrh<0$ and are physically forbidden. We thus impose
\begin{equation}
\TBBN<\Trh\leq\Trhmax,
\qquad
\TBBN=10~{\rm MeV},
\label{eq:Trange}
\end{equation}
which lies above the lower bounds $\Trh\gtrsim2$--$5$~MeV from BBN and the CMB, depending on the decay channel~\cite{deSalas2015,Hasegawa2019}. The physical interval therefore spans $-2<\log_{10}(\Trh/{\rm GeV})\lesssim15.5$.

\begin{figure}[tbp]
\centering
\includegraphics[width=\columnwidth]{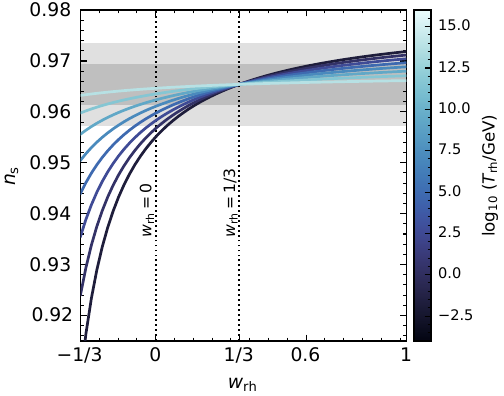}
\caption{Scalar spectral index $\ns$ as a function of $\wrh$ for representative reheating temperatures, computed with the modified \texttt{CLASS} used for the inference at fixed $\alpha=1$ and $A_s$. The vertical lines mark the matter-like value expected for oscillations in the quadratic minimum, $\wrh=0$, and the radiation-like value $\wrh=1/3$, at which the dependence on $\Trh$ disappears. The dark and light grey bands show the $1\sigma$ and $2\sigma$ Planck-only constraint of our previous analysis~\cite{Carrion2026}, $\ns=0.9654\pm0.0041$, obtained with $\Nstar$ free.}
\label{fig:ns-wrh-prior}
\end{figure}

The dependence of $\ns$ on $\wrh$ for different reheating temperatures, at fixed $\alpha=1$, is illustrated in \cref{fig:ns-wrh-prior}. Lowering $\Trh$ decreases $\ns$ for $\wrh<1/3$ and increases it for $\wrh>1/3$, while all curves meet at $\wrh=1/3$, as follows from \cref{eq:DeltaN}.

\begin{figure}[tbp]
\centering
\includegraphics[width=\columnwidth]{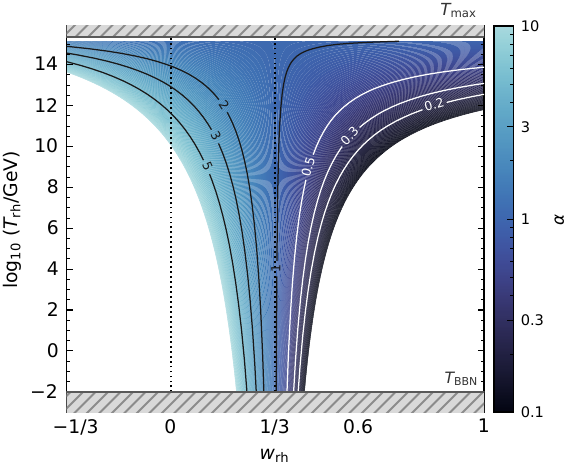}
\caption{Value of $\alpha$ for which the matched solution reproduces $\ns=0.9654$, the Planck-only value shown in \cref{fig:ns-wrh-prior}, at fixed $A_s=2.1\times10^{-9}$, in the $(\wrh,\log_{10}(\Trh/{\rm GeV}))$ plane. Each point is a full computation with the modified \texttt{CLASS} on a grid of thirteen values $0.1\le\alpha\le10$, interpolated in $\log\alpha$. White regions admit no solution in this range. Dotted lines mark $\wrh=0$ and $1/3$. The hatched bands show $\Trh<\TBBN=10$~MeV and $\Trh>\Trhmax$ (the latter for the smallest $\Trhmax$ in the $\alpha$ range). Along $\wrh=1/3$ the required $\alpha$ is independent of $\Trh$, as follows from \cref{eq:DeltaN}.}
\label{fig:reheating-impact}
\end{figure}

When $\alpha$ is allowed to vary, the same value of $\ns$ can be obtained either by changing the shape of the potential or by changing the reheating history. \Cref{fig:reheating-impact} illustrates this interplay: it shows, in the $(\wrh,\log_{10}\Trh)$ plane, the value of $\alpha$ for which the matched solution reproduces a fixed value of $\ns$.

Below $\wrh=1/3$, where $\Nstar<N_{\rm inst}$, reproducing the same $\ns$ requires $\alpha>1$, because a larger $\alpha$ raises both $\ns$ at fixed $\Nstar$ and $N_{\rm inst}$ itself. Above $\wrh=1/3$ the stiff branch allows $\alpha<1$. These scans use the same \texttt{CLASS} implementation as the inference and serve only to illustrate how the matching depends on the reheating parameters.

\section{Bayesian analysis and numerical implementation}
\label{sec:method}

\subsection{Numerical implementation}
\label{sec:numerics}

Our analysis builds on the pipeline of~\cite{Carrion2026}. We adopt the pivot scale $k_*=0.05\,{\rm Mpc}^{-1}$, set $a_0=1$, and use $T_0=2.7255\,{\rm K}$. The $\alpha$-Starobinsky potential is implemented in a modified version of the Boltzmann solver \texttt{CLASS}~\cite{Lesgourgues2011,CLASS2011}\footnote{\url{https://github.com/karimpsi22/class_alphastaro.git}}, with the reheating matching of \cref{eq:matching-numerical,eq:matching-condition} incorporated into the inflationary module. Since \texttt{CLASS} sets the unreduced Planck mass to unity, $\mpl=1/\sqrt{8\pi}$ in code units, and a tabulated value of $V_0$ corresponds to $V_0(\sqrt{8\pi}\,\mpl)^4$ in physical units. The same conversion is used when expressing the inflaton mass in GeV.

Unlike in the previous analysis, $\Nstar$ is not supplied as an input. For each proposed parameter set, \texttt{CLASS} integrates the inflationary background, solves \cref{eq:matching-condition} for $(\phi_*,\Nstar)$, and evolves the scalar and tensor perturbations mode by mode on the exact background to obtain the primordial spectra. These are propagated through the Boltzmann hierarchy to yield the CMB angular power spectra $C_\ell^{\rm TT}$, $C_\ell^{\rm TE}$ and $C_\ell^{\rm EE}$, together with the lensing prediction where required. The quantities $\Nstar$, $\phi_*$, $\Nrh$, $\rho_{\rm rh}$, $\Trhmax$ and $\ln R_{\rm rad}$ are derived. Configurations that violate \cref{eq:Trange}, or for which inflation does not end, are rejected as a recoverable computation error. The semi-analytical relations of \cref{eq:phiend,eq:NSR,eq:matching,eq:rhoend} were used for validation and interpretation only. 

Bayesian inference is performed with \texttt{Cobaya}~\cite{Cobaya2021}\footnote{\url{https://github.com/CobayaSampler/cobaya}}. We sample with \texttt{pocoMC}~\cite{Karamanis2022pocoMC,Karamanis2022MNRAS,KaramanisSeljak2024}\footnote{\url{https://github.com/minaskar/pocomc}}, a preconditioned sequential Monte Carlo sampler that moves a population of particles through a sequence of tempered posteriors, $p_\beta\propto\mathcal L^\beta\pi$, from the prior ($\beta=0$) to the posterior ($\beta=1$). A run ends when the inverse temperature reaches $\beta=1$ and the effective sample size reaches its target, and all our runs meet this criterion. Posterior summaries use the importance weights of the final sample and are computed with \texttt{GetDist}~\cite{LewisGetDist2025}\footnote{\url{https://github.com/cmbant/getdist}}. The sampler settings, diagnostics and tests with more particles are given in \cref{app:sampler-validation}.

\begin{figure*}[t]
\centering
\resizebox{\textwidth}{!}{%
\begin{tikzpicture}[
    font=\sffamily,
    every node/.append style={inner xsep=3pt},
    >={Stealth[length=2mm,width=1.5mm]},
    base/.style={draw, fill=white, rounded corners=2.5pt, line width=0.7pt,
                 align=center, text width=21mm, inner sep=3pt,
                 minimum height=12mm, font=\tiny},
    th/.style ={base, draw=cTh},
    cl/.style ={base, draw=cCl},
    inf/.style={base, draw=cIn},
    dat/.style={base, draw=cDa, dash pattern=on 2pt off 1.7pt, fill=cDa!6},
    flow/.style={->, line width=0.8pt, draw=black!62},
    link/.style={-,  line width=0.8pt, draw=black!62},
    dot/.style={circle, fill=black!62, inner sep=0pt, minimum size=2.4pt},
    panel/.style={rounded corners=5pt, line width=0.55pt},
    stg/.style={font=\tiny\bfseries\sffamily, inner sep=0pt}
  ]
  \node[th]  (pars)  at ( 0.0, 0.0) {{\color{cTh}\bfseries Sampled\\ parameters}\\[2pt]
        $V_0,\ \alpha,\ w_{\rm rh},\ T_{\rm rh}$\\[1pt]
        late-time, nuisance};
  \node[th]  (match) at ( 2.7, 0.0) {{\color{cTh}\bfseries Pivot matching}\\[2pt]
        root search in $\phi_*$:\\[1pt] $N^{\rm bg}_*=N^{\rm match}_*$};
  \node[cl]  (prim)  at ( 5.7, 1.15) {{\color{cCl}\bfseries Primordial spectra}\\[2pt]
        $\mathcal{P}_{\mathcal{R}}(k),\ \mathcal{P}_{T}(k)$\\[1pt]
        {\color{black!55}mode by mode}};
  \node[cl]  (cmb)   at ( 8.6, 1.15){{\color{cCl}\bfseries CMB spectra}\\[2pt]
        $C_\ell^{TT},\,C_\ell^{TE},\,C_\ell^{EE}$, $C_\ell^{\phi\phi}$};
  \node[cl]  (bao)   at ( 8.6,-1.15){{\color{cCl}\bfseries Background}\\[2pt]
        $D_M/r_{\rm d},\ D_H/r_{\rm d}$};
  \node[inf] (like)  at (11.7, 0.0) {{\color{cIn}\bfseries Joint\\ likelihood}\\[2pt]
        $\ln\mathcal{L}(\mathcal{D}\,|\,\Theta)$};
  \node[inf] (poco)  at (14.4, 0.0) {{\color{cIn}\bfseries\texttt{pocoMC}}\\[2pt]
        sampler};
  \node[inf] (post)  at (17.1, 0.0) {{\color{cIn}\bfseries Posteriors}\\[2pt]
        $p(\Theta\,|\,\mathcal{D})$};
  \node[dat] (data) at (11.7,-2.95) {{\color{cDa}\bfseries Data}\\[2pt]
    \begin{tabular}{@{}l@{}}
    $\bullet$\ Planck\\
    $\bullet$\ ACT\\
    $\bullet$\ DESI
    \end{tabular}};
  \draw[flow] (pars)  -- (match);
  \draw[flow] (match.east) -- ++(0.35,0) |- (prim.west);
  \draw[flow] (match.east) -- ++(0.35,0) |- (bao.west);
  \draw[flow] (prim)  -- (cmb);
  \coordinate (jn) at (10.15, 0);
  \draw[link] (cmb.east) -- (10.15, 1.15) -- (jn);
  \draw[link] (bao.east) -- (10.15,-1.15) -- (jn);
  \draw[flow] (jn) -- (like.west);
  \node[dot] at (jn) {};
  \draw[flow] (data) -- (like);
  \draw[flow] (like) -- (poco);
  \draw[flow] (poco) -- (post);
  \draw[flow, dashed] (poco.north) -- ++(0,1.95) -| (pars.north);
  \coordinate (tt) at (0,1.95);   \coordinate (bb) at (0,-1.85);
  \begin{scope}[on background layer]
    \node[panel, fill=cTh!7, draw=cTh!30, inner sep=2.8mm,
          fit=(pars)(match)(pars|-tt)(pars|-bb)] (pTh) {};
    \node[panel, fill=cCl!7, draw=cCl!30, inner sep=2.8mm,
          fit=(prim)(cmb)(bao)(prim|-tt)(prim|-bb)] (pCl) {};
    \node[panel, fill=cIn!7, draw=cIn!30, inner sep=2.8mm,
          fit=(like)(poco)(post)(like|-tt)(like|-bb)] (pIn) {};
  \end{scope}
  \node[stg, text=cTh!80!black, anchor=north] at ($(pTh.south)+(0,-1.4mm)$)
        {I. MODEL AND REHEATING};
  \node[stg, text=cCl!65!black, anchor=north] at ($(pCl.south)+(0,-1.4mm)$)
        {II. BOLTZMANN SOLVER (\texttt{CLASS})};
  \node[stg, text=cIn!80!black, anchor=north] at ($(pIn.south)+(0,-1.4mm)$)
        {III. INFERENCE (\texttt{Cobaya})};
\end{tikzpicture}}
\caption{Inference pipeline. \textbf{(I)}~For each sampled point, \texttt{CLASS} integrates the inflationary background and solves the matching condition, \cref{eq:matching-condition}, for $\phi_*$ and $\Nstar$, and points that violate the conditions of \cref{tab:priors} are rejected (zero likelihood). \textbf{(II)}~The perturbations are evolved on the exact background, and the primordial spectra are propagated to the CMB spectra, while the late-time background gives the BAO distances. \textbf{(III)}~The predictions are confronted with the Planck, ACT and DESI likelihoods, and \texttt{pocoMC} proposes new points (dashed arrow) until the target effective sample size is reached.}
\label{fig:inference-pipeline}
\end{figure*}
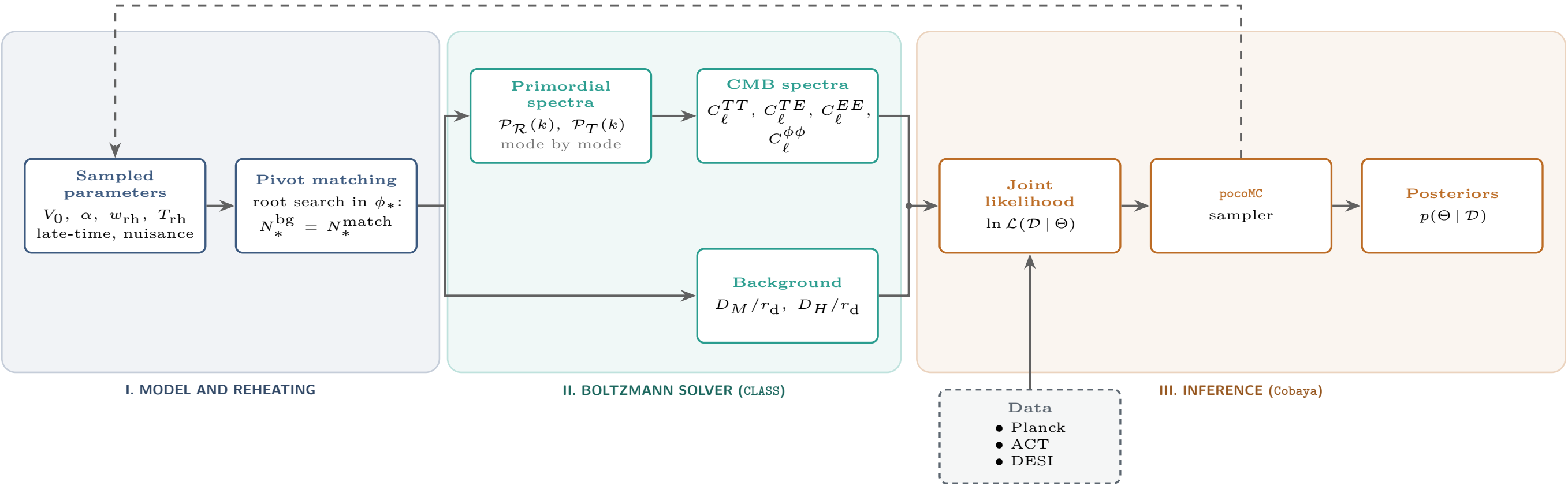

\subsection{Data and likelihoods}
\label{sec:data}

We combine primary-CMB information from Planck and ACT with BAO measurements from DESI DR2. Following the community convention, we denote by ``P-ACT'' the joint Planck and ACT primary-CMB combination with the multipole cuts defined below. CMB lensing is included only in the Planck-only combinations. Since our combinations involving ACT do not include it, they do not coincide with the ACT ``P-ACT-LB'' combination, which includes CMB lensing and BAO. We denote our combinations explicitly as ``P-ACT+DESI''.

For the Planck-only combinations we use the Planck 2018 low-$\ell$ TT and EE likelihoods~\cite{Planck2018V}, the NPIPE (PR4) CamSpec high-$\ell$ TT/TE/EE likelihood~\cite{Rosenberg2022}, and the Planck PR4 lensing likelihood~\cite{Carron2022}\footnote{\url{https://github.com/carronj/planck_PR4_lensing}}. For ACT we use the foreground-marginalized ACT DR6 primary-CMB likelihood\footnote{\url{https://github.com/ACTCollaboration/DR6-ACT-lite}} over its validated multipole range, approximately $600\leq\ell\leq6500$ in TT, TE and EE~\cite{Louis2025}.

Since Planck and ACT observe overlapping regions of the sky, P-ACT uses the multipole cuts supplied with the ACT DR6 likelihood to avoid counting the same CMB modes twice: the Planck 2018 (PR3) Plik-lite likelihood is restricted to $\ell\le1000$ in TT and $\ell\le600$ in TE and EE, and ACT supplies the smaller scales. All combinations with ACT include the Planck low-$\ell$ EE likelihood in its SRoll2 version~\cite{Pagano2020}. The Planck data also differ between the Planck-only and P-ACT combinations, at high $\ell$ (CamSpec PR4 versus cut Plik-lite PR3) and in the low-$\ell$ EE likelihood (SimAll versus SRoll2), so differences between them reflect both the added ACT data and these choices.

The late-time expansion history is constrained by the DESI DR2 BAO likelihood~\cite{DESIDR2}, which measures the BAO scale relative to the sound horizon. The resulting six combinations are listed in \cref{tab:data-combinations}. The first three define the CMB-only baseline, and the remaining three isolate the effect of adding DESI.\footnote{Within $\Lambda$CDM, ACT alone and DESI DR2 are in tension at more than $3\sigma$, a tension that is reduced when ACT is combined with Planck~\cite{GarciaQuintero2025}, so results for ACT+DESI should be read with this caveat.} The pipeline is summarized in \cref{fig:inference-pipeline}.

\begin{table*}[t]
\centering
\footnotesize
\renewcommand{\arraystretch}{1.22}
\begin{tabular}{@{}lp{0.83\textwidth}@{}}
\toprule
Combination & Likelihood modules in \texttt{Cobaya} \\
\midrule
${\cal D}_1$: Planck &
\path{planck_2018_lowl.TT}, \path{planck_2018_lowl.EE}, \path{planck_NPIPE_highl_CamSpec.TTTEEE}, \path{planckpr4lensing} \\
${\cal D}_2$: ACT &
\path{act_dr6_cmbonly}, \path{planck_2018_lowl.EE_sroll2} \\
${\cal D}_3$: P-ACT &
\path{act_dr6_cmbonly}, \path{act_dr6_cmbonly.PlanckActCut} (Plik-lite PR3, TT $\ell\le1000$, TE/EE $\ell\le600$), \path{planck_2018_lowl.TT}, \path{planck_2018_lowl.EE_sroll2} \\
${\cal D}_4$: Planck+DESI & ${\cal D}_1$ $+$ \path{bao.desi_dr2} \\
${\cal D}_5$: ACT+DESI & ${\cal D}_2$ $+$ \path{bao.desi_dr2} \\
${\cal D}_6$: P-ACT+DESI & ${\cal D}_3$ $+$ \path{bao.desi_dr2} \\
\bottomrule
\end{tabular}
\caption{Data combinations used in the analysis and the corresponding likelihood modules.}
\label{tab:data-combinations}
\end{table*}

\subsection{Sampled parameters and priors}
\label{sec:priors}

The sampled parameters are
\begin{equation}
\begin{split}
\boldsymbol{\theta}
={}&
\underbrace{\{\log_{10}V_0,\,\log_{10}\alpha,\,\wrh,\,\log_{10}(\Trh/{\rm GeV})\}}_{\text{inflation and reheating}}
\\
&\cup\;
\underbrace{\{100\,\theta_s,\,\Omega_b h^2,\,\Omega_c h^2,\,\tau_{\rm reio}\}}_{\text{late-time cosmology}}
\\
&\cup\;
\text{nuisance parameters},
\end{split}
\label{eq:theta}
\end{equation}
where $\theta_s$ is the angular size of the sound horizon at recombination, $\Omega_b h^2$ and $\Omega_c h^2$ are the physical baryon and cold dark matter densities, and $\tau_{\rm reio}$ is the optical depth to reionization. For the canonical analysis $\alpha=1$ is fixed. The priors, summarized in \cref{tab:priors}, are common to the six data combinations. We assume one massive neutrino with $\sum m_\nu = 0.06\,{\rm eV}$ and $N_{\rm eff}=3.044$, and fix $\grh=\gsrh=106.75$.

\begin{table}[tbp]
\centering
\footnotesize
\renewcommand{\arraystretch}{1.22}
\setlength{\tabcolsep}{5pt}
\begin{tabular}{@{}lcc@{}}
\toprule
Parameter & Prior & Range \\
\midrule
\multicolumn{3}{@{}l}{\textbf{Inflation and reheating}}\\
$\log_{10}V_0$ & Uniform & $[-18,-10]$ \\
$\log_{10}\alpha$ & Uniform & $[-2,2]$ \\
$\wrh$ (permissive) & Uniform & $(-1/3,\,1]$ \\
$\wrh$ (restrictive) & Uniform & $(-1/3,\,1/3]$ \\
$\log_{10}(\Trh/{\rm GeV})$ & Uniform & $[-2,16]$ \\
\addlinespace[2pt]
\multicolumn{3}{@{}l}{\textbf{Late-time cosmology}}\\
$100\,\theta_s$ & Uniform & $[1.02,1.06]$ \\
$\Omega_b h^2$ & Uniform & $[0.019,0.025]$ \\
$\Omega_c h^2$ & Uniform & $[0.08,0.16]$ \\
$\tau_{\rm reio}$ & Uniform & $[0.03,0.15]$ \\
\addlinespace[2pt]
\multicolumn{3}{@{}l}{\textbf{Nuisance}}\\
$A_{\rm ACT}$ & Uniform $\times\,\mathcal{N}(1,0.003)$ & $[0.97,1.03]$ \\
others & default likelihood settings & -- \\
\addlinespace[2pt]
\multicolumn{3}{@{}l}{\textbf{Additional conditions}}\\
$\Trh$ & & $\Trh\le\Trhmax$ \\
$\Nstar$ & & $[20,90]$ \\
\bottomrule
\end{tabular}
\caption{Sampled parameters, their priors and the additional conditions, common to the six data combinations. Configurations violating these conditions are rejected, so the induced prior on $(\wrh,\Trh)$ is the product of the uniform priors truncated by these conditions.}
\label{tab:priors}
\end{table}

The prior ranges of the late-time parameters are narrower than the broad boxes commonly used in Metropolis--Hastings analyses. This choice is motivated by the sampling strategy: \texttt{pocoMC} draws its initial particles from the prior, so the prior volume is also the initial search volume. With the broad default ranges of \texttt{Cobaya}, most of these initial points do not give a computable cosmology, and the first tempering steps are wasted. As a check, for the $\alpha$-Starobinsky model without reheating and Planck 2018 data, \texttt{pocoMC} with these narrower ranges recovers the posteriors of a Metropolis--Hastings run with broad priors. The adopted ranges lie at least $4\sigma$ from the Planck posterior mean in $\tau_{\rm reio}$ (lower edge) and more than $20\sigma$ away in the other late-time parameters, so they do not truncate the posterior. For the ACT calibration we sample $A_{\rm ACT}\in[0.97,1.03]$ while retaining the Gaussian $\mathcal{N}(1,0.003)$ prior, whose $1\sigma$ width is a tenth of the half-width of the sampled interval.

At fixed $(\alpha,\wrh,\Trh)$ the scalar amplitude scales as $A_s\propto V_0\Nstar^2/\alpha$, from \cref{eq:attractor}, with only a weak dependence of $\Nstar$ on $V_0$, so a uniform prior on $\log_{10}V_0$ is equivalent to a uniform prior on $\ln A_s$ and does not affect the induced prior on the reheating sector. Configurations with $\Trh>\Trhmax$ are rejected at the level of the likelihood. Since $\Trhmax\propto V_0^{1/4}$ ranges from about $10^{14}$ to $10^{16}$~GeV across the $\log_{10}V_0$ prior at $\alpha=1$, the induced prior on $\log_{10}(\Trh/{\rm GeV})$ is flat up to $\simeq14$ and falls to zero at $\simeq16$. Once the data fix $V_0$ through $A_s$, the cut lies at $\log_{10}(\Trhmax/{\rm GeV})\simeq15.35$. The lower edge of the $\log_{10}\alpha$ prior is set by the numerical stability of the inflationary solver, which fails for $\alpha\lesssim5\times10^{-3}$. Since the data cannot distinguish small values of $\alpha$, the posterior mass there depends on this edge.

For $\TBBN<\Trh\le\Trhmax$ and $-1/3<\wrh\le1$, \cref{eq:DeltaN} restricts $\Nstar$ to $[15.9,68.9]$ at $\alpha=1$. The rejection of $\Nstar<20$ removes only configurations with $\wrh\lesssim-0.30$ and $\Trh$ near the BBN bound, a negligible fraction of the prior volume, and the upper rejection bound, $\Nstar=90$, is never reached.

\section{Results}
\label{sec:results}

We first present the constraints for the canonical model and for free $\alpha$, obtained by sampling the reheating model directly. We then compare them with an analysis without a reheating sector, which isolates the effect of the reheating prior and also tests the direct sampling, and finally repeat the free-$\alpha$ analysis with the stiff branch forbidden.

\subsection{The canonical Starobinsky model}
\label{sec:canonical}

\begin{figure*}[t]
\centering
\includegraphics[width=0.80\textwidth]{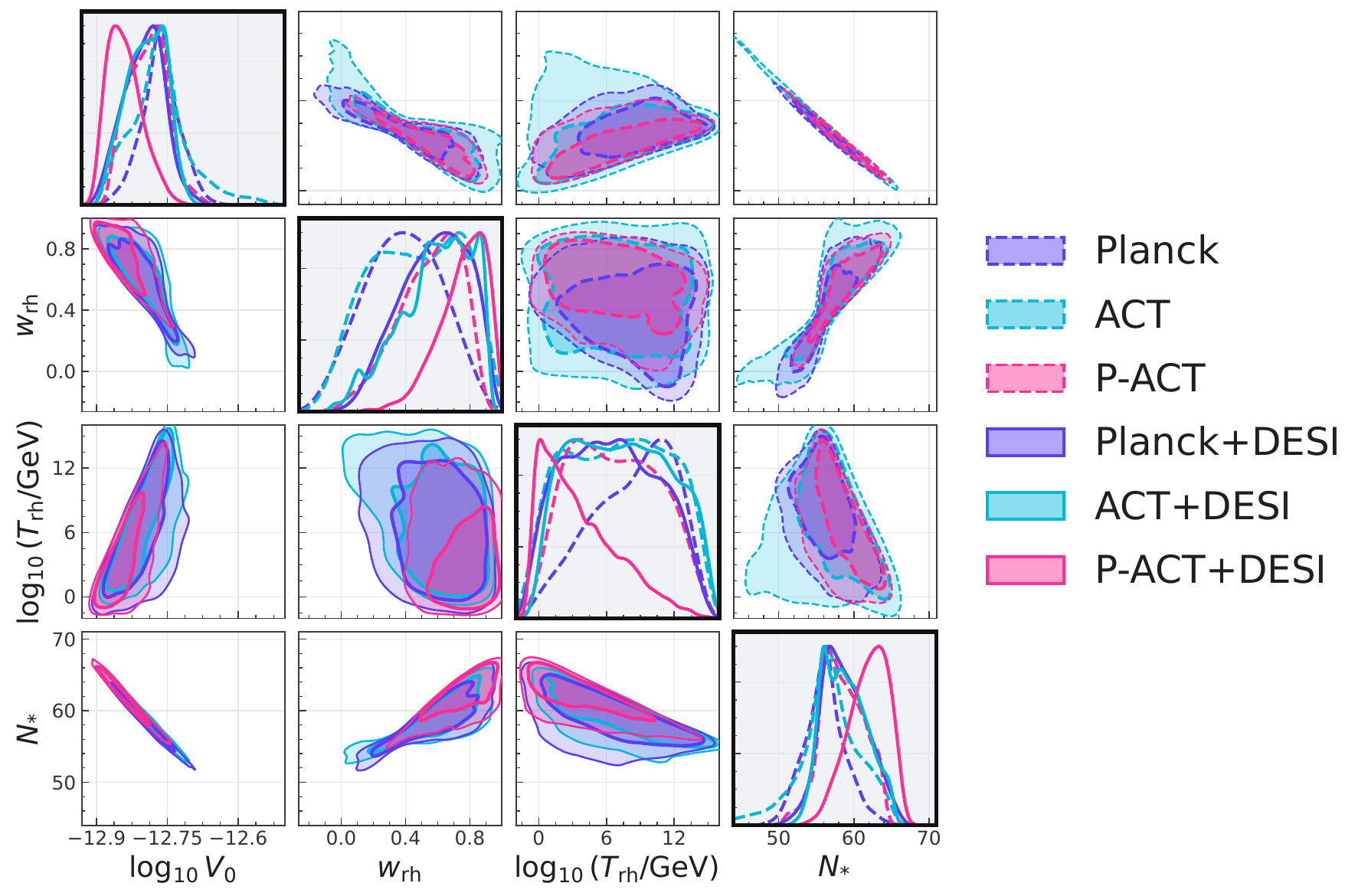}
\caption{Marginalized posteriors of $\log_{10}V_0$, $\wrh$, $\log_{10}(\Trh/{\rm GeV})$ and $\Nstar$ for the canonical model ($\alpha=1$). Dashed lines: Planck, ACT and P-ACT. Solid lines: the same combinations with DESI. Adding DESI moves the posterior toward larger $\Nstar$ and $\wrh$, most clearly for P-ACT+DESI, which also prefers lower $\Trh$.}
\label{fig:alpha1-corner}
\end{figure*}

We start with $\alpha=1$. With the shape of the potential fixed, the only freedom left is the point of the potential at which the pivot scale leaves the horizon, that is, $\Nstar$. A larger $\Nstar$ gives a larger $\ns$ and a smaller $r$, so all data combinations lie on one curve in the $(\ns,r)$ plane and differ only in where they sit along it. \Cref{tab:alpha1-results} collects the constraints and \cref{fig:alpha1-corner} shows the marginalized posteriors.

Planck alone gives $\ns=0.9656^{+0.0015}_{-0.0017}$, consistent with instantaneous reheating: $\ln R_{\rm rad}=0.4\pm2.7$. ACT alone gives similar constraints. Combining Planck with ACT, and adding DESI, raises $\ns$. DESI prefers a smaller $\Omega_{\rm m}$ and a larger $H_0$, and both are correlated with $\ns$ in the CMB fit. The shift is largest for P-ACT+DESI, with $\ns=0.9689^{+0.0012}_{-0.0016}$ and $\ln R_{\rm rad}=6.5^{+2.5}_{-3.1}$, so $\Nstar$ lies about $6.5$ $e$-folds above its instantaneous-reheating value.

\begin{table*}[t]
\centering
\footnotesize
\setlength{\tabcolsep}{2.4pt}
\renewcommand{\arraystretch}{1.45}
\begin{tabular}{lcccccc}
\toprule
\textbf{Parameter} & \textbf{Planck} & \textbf{ACT} & \textbf{P-ACT} &
\textbf{Planck+DESI} & \textbf{ACT+DESI} & \textbf{P-ACT+DESI} \\
\midrule
\multicolumn{7}{l}{\textit{Sampled}} \\
$\log_{10} V_0$ & $-12.77\pm0.04$ & $-12.77\pm0.06$ & $-12.79^{+0.04}_{-0.05}$ & $-12.80^{+0.04}_{-0.05}$ & $-12.79^{+0.04}_{-0.05}$ & $-12.85^{+0.04}_{-0.03}$ \\
$\wrh$ & $0.39\pm0.25$ & $0.5\pm0.3$ & $0.58^{+0.18}_{-0.24}$ & $0.59^{+0.21}_{-0.24}$ & $0.61^{+0.21}_{-0.27}$ & $0.77^{+0.13}_{-0.18}$ \\
$\log_{10}(\Trh/\mathrm{GeV})$ & $9^{+3}_{-5}$ & $7\pm5$ & $7^{+5}_{-4}$ & $6^{+5}_{-4}$ & $7\pm5$ & $<10.3$ \\
$100\,\theta_\mathrm{s}$ & $1.04176\pm0.00023$ & $1.04151^{+0.00028}_{-0.00029}$ & $1.04161\pm0.00025$ & $1.04191^{+0.00021}_{-0.00022}$ & $1.04185^{+0.00027}_{-0.00025}$ & $1.04176\pm0.00025$ \\
$\Omega_\mathrm{b} h^2$ & $0.02221\pm0.00013$ & $0.02256\pm0.00015$ & $0.02244\pm0.00011$ & $0.02232\pm0.00011$ & $0.02271\pm0.00014$ & $0.02257\pm0.00010$ \\
$\Omega_\mathrm{c} h^2$ & $0.1193\pm0.0009$ & $0.1240^{+0.0021}_{-0.0019}$ & $0.1203\pm0.0009$ & $0.1177\pm0.0006$ & $0.1177\pm0.0007$ & $0.1180\pm0.0006$ \\
$\tau_\mathrm{reio}$ & $0.054\pm0.006$ & $0.056^{+0.006}_{-0.005}$ & $0.059\pm0.006$ & $0.059\pm0.006$ & $0.061\pm0.006$ & $0.061^{+0.006}_{-0.005}$ \\
\midrule
\multicolumn{7}{l}{\textit{Derived}} \\
$\Nstar$ & $56.1\pm2.7$ & $57\pm4$ & $58.3^{+3.3}_{-2.6}$ & $58.5^{+3.5}_{-2.7}$ & $58.6^{+3.4}_{-2.8}$ & $62.2^{+2.5}_{-3.1}$ \\
$\ln R_{\rm rad}\simeq\Nstar-N_{\rm inst}$ & $0.4\pm2.7$ & $1\pm4$ & $2.6^{+3.4}_{-2.6}$ & $2.8^{+3.5}_{-2.7}$ & $2.9^{+3.5}_{-2.8}$ & $6.5^{+2.5}_{-3.1}$ \\
$\Nrh$ & $15^{+9}_{-8}$ & $17^{+11}_{-10}$ & $17^{+7}_{-9}$ & $17^{+8}_{-9}$ & $16\pm9$ & $22^{+5}_{-8}$ \\
$\ns$ & $0.9656^{+0.0015}_{-0.0017}$ & $0.9660^{+0.0023}_{-0.0022}$ & $0.9668^{+0.0017}_{-0.0015}$ & $0.9670^{+0.0018}_{-0.0015}$ & $0.9670^{+0.0018}_{-0.0016}$ & $0.9689^{+0.0012}_{-0.0016}$ \\
$r\ [10^{-3}]$ & $3.41^{+0.34}_{-0.29}$ & $3.3\pm0.4$ & $3.17^{+0.28}_{-0.32}$ & $3.15^{+0.29}_{-0.33}$ & $3.1\pm0.3$ & $2.80^{+0.28}_{-0.21}$ \\
$\ln(10^{10} A_\mathrm{s})$ & $3.040^{+0.012}_{-0.013}$ & $3.053^{+0.013}_{-0.012}$ & $3.055\pm0.012$ & $3.048\pm0.012$ & $3.060\pm0.014$ & $3.056^{+0.013}_{-0.012}$ \\
$H_0$ & $67.4\pm0.4$ & $66.0^{+0.7}_{-0.8}$ & $67.3\pm0.4$ & $68.15\pm0.26$ & $68.49^{+0.30}_{-0.29}$ & $68.23^{+0.25}_{-0.26}$ \\
$\Omega_\mathrm{m}$ & $0.312\pm0.005$ & $0.337^{+0.013}_{-0.011}$ & $0.317^{+0.006}_{-0.005}$ & $0.303\pm0.003$ & $0.301\pm0.004$ & $0.303\pm0.003$ \\
$\sigma_8$ & $0.807\pm0.005$ & $0.826\pm0.007$ & $0.816^{+0.005}_{-0.006}$ & $0.805\pm0.005$ & $0.809\pm0.006$ & $0.809^{+0.006}_{-0.005}$ \\
\bottomrule
\end{tabular}
\caption{Marginalized constraints for the canonical model ($\alpha=1$), permissive prior $-1/3<\wrh\le1$. Central values are medians and two-sided uncertainties are 68\% equal-tail credible intervals. For asymmetric intervals the rounding is set by the smaller uncertainty. When the posterior of $\log_{10}\Trh$ reaches the lower prior edge, we quote the one-sided 95\% upper bound instead.}
\label{tab:alpha1-results}
\end{table*}

A useful rule of thumb follows from $\ns\simeq1-2/\Nstar$ in \cref{eq:attractor}: each $e$-fold of $\ln R_{\rm rad}$ shifts $\ns$ by $2/\Nstar^2\simeq6.5\times10^{-4}$ at $\alpha=1$, so that $\sigma(\ln R_{\rm rad})\simeq\frac12\Nstar^2\sigma(\ns)$. With $\sigma(\ns)\simeq0.0016$ this gives $2.5$ $e$-folds, which is the width we find for both Planck and P-ACT+DESI. The reheating constraint at fixed $\alpha$ is therefore set entirely by the precision on $\ns$. The same estimate reconciles our numbers with the range $\ns=0.958$--$0.963$ found at $\alpha=1$ for matter-like reheating with $10^2\lesssim\Trh/{\rm GeV}\lesssim10^{10}$~\cite{Ellis2025}: \cref{eq:DeltaN} gives $\ln R_{\rm rad}\simeq-4.9$ and $\ns\simeq0.962$ for $\wrh=0$ and $\Trh=10^9$~GeV, inside that range, while our permissive prior lets the data choose $\wrh$.

\begin{figure*}[t]
\centering
\includegraphics[width=0.72\textwidth]{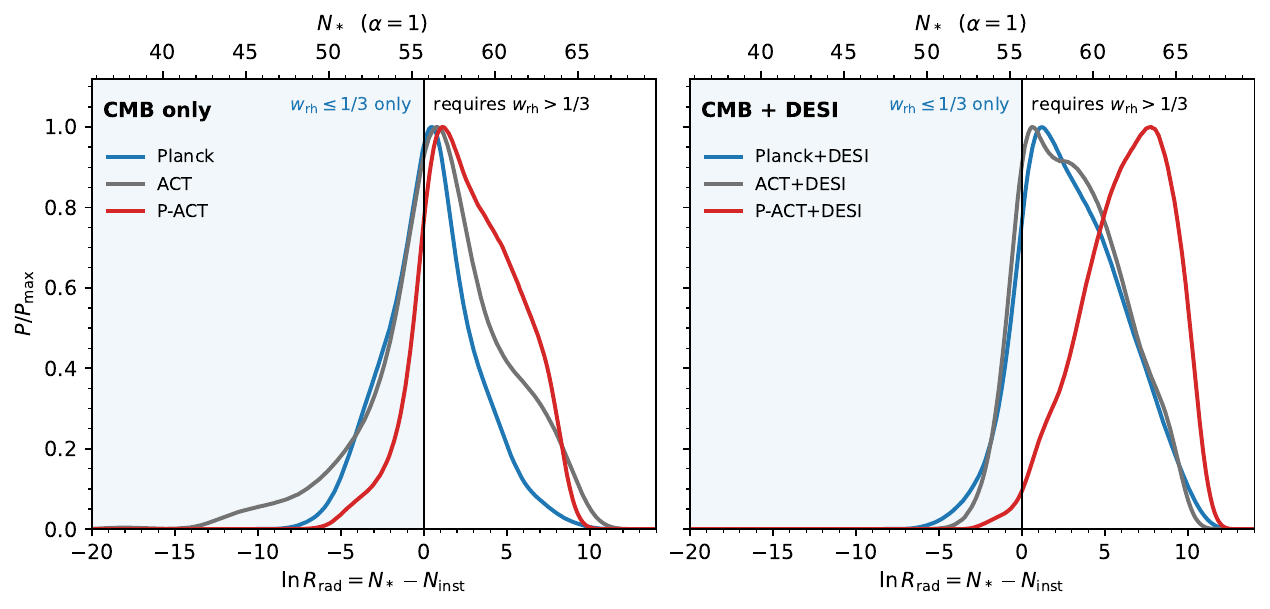}
\caption{Posterior of $\ln R_{\rm rad}\simeq\Nstar-N_{\rm inst}$ for the canonical model, CMB-only combinations (left) and with DESI (right), each normalized to unit peak. The shaded region, $\ln R_{\rm rad}\le0$, is the only one accessible with $\wrh\le1/3$, and positive values require a stiff phase. The upper axis gives $\Nstar$.}
\label{fig:lnR}
\end{figure*}

\Cref{fig:lnR} shows what this means in terms of the quantity the data actually constrain. Planck stays centered at $\ln R_{\rm rad}=0$, where, as we show later, the prior induced by the reheating sector peaks. With DESI, and above all with P-ACT+DESI, the posterior moves to positive values, which only a stiff phase can produce.

\begin{figure}[tbp]
\centering
\includegraphics[width=\columnwidth]{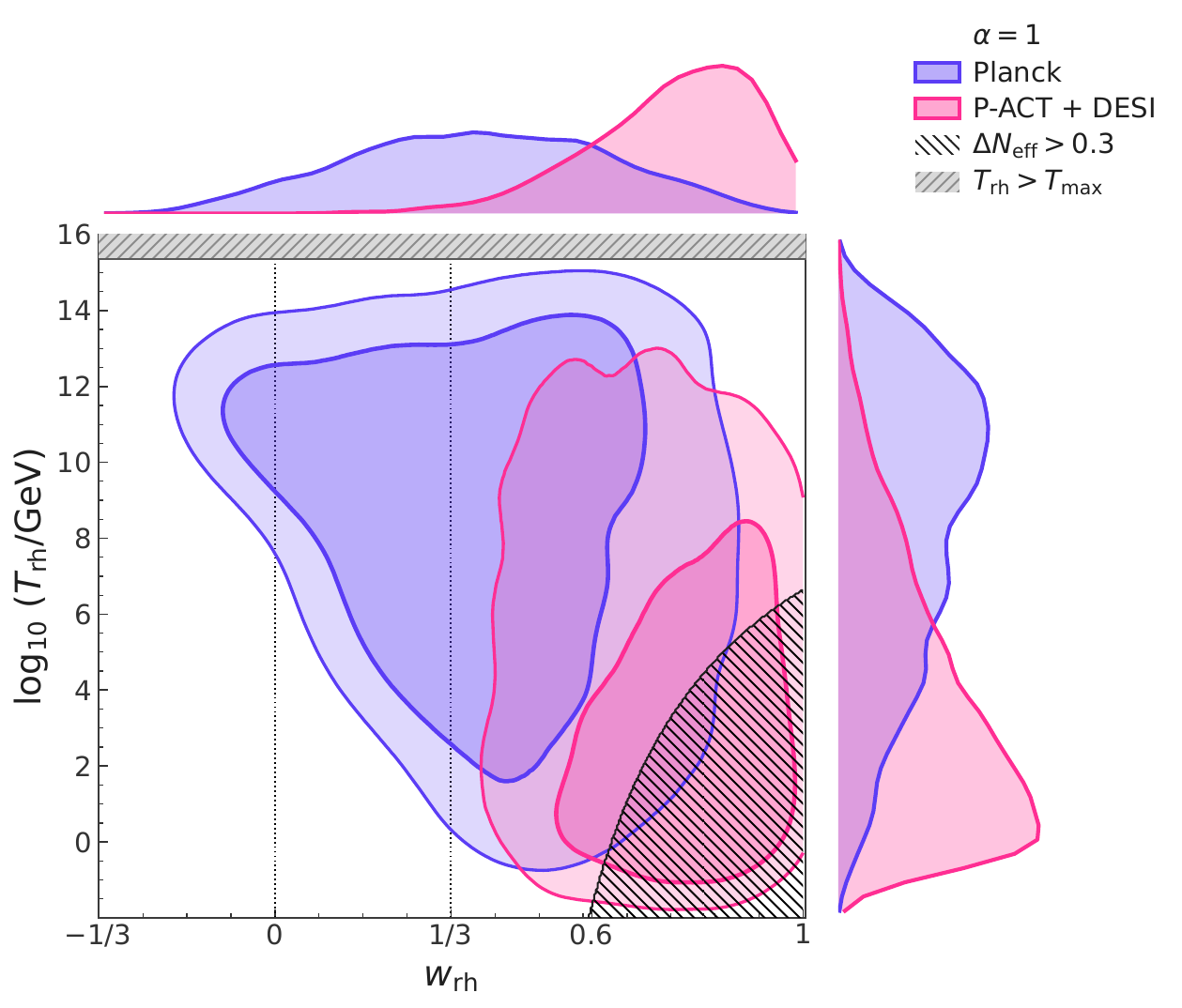}
\caption{Joint posterior in the $(\wrh,\log_{10}(\Trh/{\rm GeV}))$ plane for the canonical model, for Planck (violet) and P-ACT+DESI (magenta), with $68\%$ and $95\%$ contours and the marginal distributions on the upper and right axes. Dotted lines mark $\wrh=0$ and $1/3$. The grey hatched band is $\Trh>\Trhmax$, and the black hatched region is excluded by the approximate gravitational-wave bound $\Delta N_{\rm eff}<0.3$, computed for $r=3\times10^{-3}$. The region is only indicative, and the excluded fractions quoted in the text use the value of $r$ of each sample.}
\label{fig:plane}
\end{figure}

In terms of the original parameters, Planck leaves the equation of state broad, $\wrh=0.39\pm0.25$, with room for both the matter-like value $\wrh=0$ and the stiff branch. As the preferred $\ns$ grows, the weight moves to the stiff side. For P-ACT+DESI we find $\wrh=0.77^{+0.13}_{-0.18}$. Since this posterior runs into the prior edge at $\wrh=1$, its lower tail is the more useful number: $95\%$ of the posterior lies above $\wrh=0.47$. The reheating temperature follows the same line in the opposite direction: once $\wrh>1/3$, a lower $\Trh$ means a longer stiff phase and a larger $\Nstar$, and P-ACT+DESI gives $\log_{10}(\Trh/{\rm GeV})<10.3$ at 95\%, although, like all our temperature constraints, this bound is largely prior-dominated. \Cref{fig:plane} shows both posteriors in the $(\wrh,\Trh)$ plane, elongated along the direction of constant $\ln R_{\rm rad}$ given by \cref{eq:DeltaN}, together with the region excluded by the gravitational-wave bound.

\subsection{Free \texorpdfstring{$\alpha$}{alpha}}
\label{sec:alpha-general}

\begin{figure*}[t]
\centering
\includegraphics[width=0.85\textwidth]{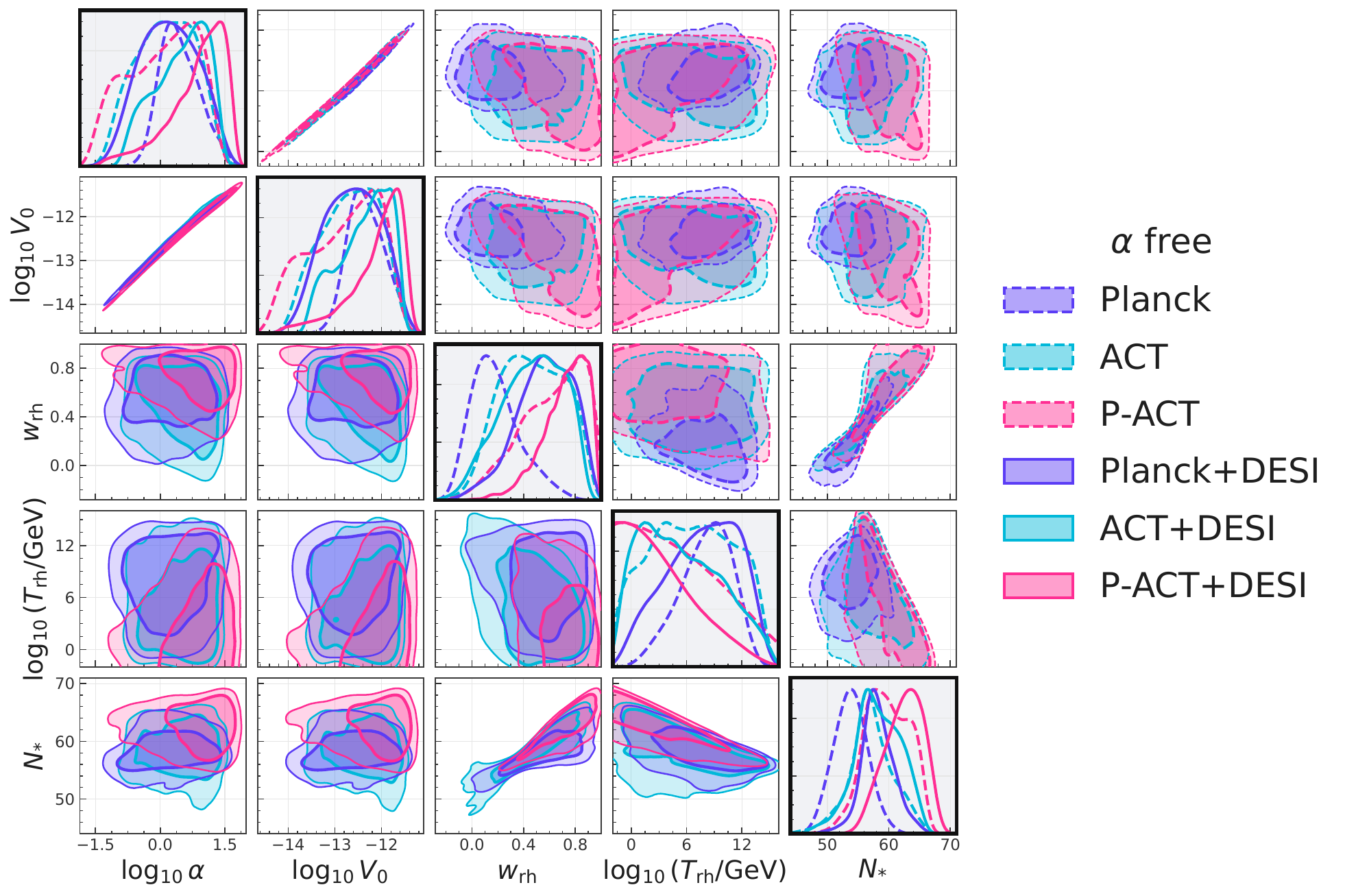}
\caption{Joint marginalized posteriors for the generalized model, showing $\log_{10}\alpha$, $\log_{10}V_0$, $\wrh$, $\log_{10}(\Trh/{\rm GeV})$ and $\Nstar$. Dashed curves and contours correspond to Planck, ACT and P-ACT, and solid contours to the combinations including DESI. Freeing $\alpha$ introduces a degeneracy between $\alpha$ and $\Nstar$, since both raise $\ns$.}
\label{fig:alpha-free-corner}
\end{figure*}

Letting $\alpha$ vary gives the model a second way to raise $\ns$: instead of increasing $\Nstar$, it can change the shape of the potential. \Cref{fig:alpha-free-corner} and \cref{tab:alpha-free-results} show the results. The extra freedom mostly shows up in $r$, while DESI still pushes $\ns$ up.

\begin{table*}[t]
\centering
\footnotesize
\setlength{\tabcolsep}{2.4pt}
\renewcommand{\arraystretch}{1.45}
\begin{tabular}{lcccccc}
\toprule
\textbf{Parameter} & \textbf{Planck} & \textbf{ACT} & \textbf{P-ACT} &
\textbf{Planck+DESI} & \textbf{ACT+DESI} & \textbf{P-ACT+DESI} \\
\midrule
\multicolumn{7}{l}{\textit{Sampled}} \\
$\log_{10}\alpha$ & $0.4^{+0.5}_{-0.4}$ & $0.1^{+0.7}_{-0.8}$ & $0.1^{+0.8}_{-1.1}$ & $0.2\pm0.7$ & $0.6^{+0.5}_{-0.8}$ & $1.0^{+0.5}_{-0.8}$ \\
$\log_{10} V_0$ & $-12.4^{+0.5}_{-0.4}$ & $-12.6^{+0.6}_{-0.7}$ & $-12.7^{+0.7}_{-1.0}$ & $-12.6\pm0.7$ & $-12.3^{+0.5}_{-0.7}$ & $-11.9^{+0.4}_{-0.7}$ \\
$\wrh$ & $0.16^{+0.22}_{-0.16}$ & $0.47^{+0.28}_{-0.26}$ & $0.65^{+0.22}_{-0.30}$ & $0.57^{+0.22}_{-0.23}$ & $0.49^{+0.24}_{-0.28}$ & $0.75^{+0.14}_{-0.20}$ \\
$\log_{10}(\Trh/\mathrm{GeV})$ & $8.4^{+2.6}_{-3.3}$ & $7\pm5$ & $<13.4$ & $8^{+4}_{-5}$ & $<13.1$ & $<11.0$ \\
$100\,\theta_\mathrm{s}$ & $1.04178^{+0.00020}_{-0.00019}$ & $1.04153^{+0.00028}_{-0.00029}$ & $1.04161^{+0.00025}_{-0.00024}$ & $1.04192\pm0.00021$ & $1.04187^{+0.00026}_{-0.00027}$ & $1.04179^{+0.00023}_{-0.00025}$ \\
$\Omega_\mathrm{b} h^2$ & $0.02221^{+0.00011}_{-0.00012}$ & $0.02254\pm0.00015$ & $0.02246\pm0.00010$ & $0.02232^{+0.00010}_{-0.00011}$ & $0.02271^{+0.00014}_{-0.00013}$ & $0.02256^{+0.00010}_{-0.00009}$ \\
$\Omega_\mathrm{c} h^2$ & $0.1191\pm0.0008$ & $0.1238^{+0.0019}_{-0.0020}$ & $0.1199\pm0.0009$ & $0.1177^{+0.0006}_{-0.0005}$ & $0.1177\pm0.0007$ & $0.1178\pm0.0006$ \\
$\tau_\mathrm{reio}$ & $0.054\pm0.006$ & $0.056^{+0.006}_{-0.005}$ & $0.059\pm0.006$ & $0.059^{+0.007}_{-0.006}$ & $0.062^{+0.006}_{-0.005}$ & $0.061\pm0.006$ \\
\midrule
\multicolumn{7}{l}{\textit{Derived}} \\
$\Nstar$ & $53.8\pm2.8$ & $57^{+4}_{-3}$ & $60\pm4$ & $58.0^{+2.8}_{-2.2}$ & $58^{+4}_{-3}$ & $63^{+3}_{-4}$ \\
$\ln R_{\rm rad}\simeq\Nstar-N_{\rm inst}$ & $-2.3\pm2.8$ & $1^{+4}_{-3}$ & $4\pm4$ & $2.1^{+2.8}_{-2.1}$ & $2^{+4}_{-3}$ & $6^{+3}_{-4}$ \\
$\Nrh$ & $19^{+8}_{-7}$ & $19\pm11$ & $20^{+8}_{-10}$ & $14^{+9}_{-8}$ & $21^{+9}_{-10}$ & $22^{+6}_{-8}$ \\
$\ns$ & $0.9651^{+0.0019}_{-0.0023}$ & $0.9665\pm0.0025$ & $0.9680^{+0.0017}_{-0.0021}$ & $0.9674^{+0.0019}_{-0.0022}$ & $0.9678^{+0.0022}_{-0.0024}$ & $0.9706^{+0.0016}_{-0.0020}$ \\
$r\ [10^{-3}]$ & $8^{+13}_{-5}$ & $4^{+12}_{-4}$ & $4^{+14}_{-3}$ & $5^{+14}_{-4}$ & $10^{+16}_{-8}$ & $19^{+16}_{-15}$ \\
$\ln(10^{10} A_\mathrm{s})$ & $3.039^{+0.012}_{-0.011}$ & $3.054^{+0.013}_{-0.012}$ & $3.055^{+0.013}_{-0.012}$ & $3.046^{+0.013}_{-0.012}$ & $3.060^{+0.014}_{-0.012}$ & $3.056\pm0.013$ \\
$H_0$ & $67.5\pm0.4$ & $66.1\pm0.7$ & $67.4\pm0.4$ & $68.18^{+0.24}_{-0.26}$ & $68.5\pm0.3$ & $68.29^{+0.25}_{-0.24}$ \\
$\Omega_\mathrm{m}$ & $0.312\pm0.005$ & $0.337\pm0.012$ & $0.315\pm0.005$ & $0.303\pm0.003$ & $0.301\pm0.004$ & $0.302\pm0.003$ \\
$\sigma_8$ & $0.806\pm0.005$ & $0.826\pm0.007$ & $0.815\pm0.006$ & $0.804\pm0.005$ & $0.809\pm0.006$ & $0.809\pm0.006$ \\
\bottomrule
\end{tabular}
\caption{As \cref{tab:alpha1-results}, for the generalized model with the prior $\log_{10}\alpha\rightarrow\mathcal U[-2,2]$.}
\label{tab:alpha-free-results}
\end{table*}

For Planck we find $\log_{10}\alpha=0.4^{+0.5}_{-0.4}$, $\Nstar=53.8\pm2.8$ and $\ns=0.9651^{+0.0019}_{-0.0023}$. P-ACT+DESI again moves furthest, to $\log_{10}\alpha=1.0^{+0.5}_{-0.8}$, $\Nstar=63^{+3}_{-4}$ and $\ns=0.9706^{+0.0016}_{-0.0020}$. This suggests $\alpha>1$, but the preference is weak. P-ACT and Planck+DESI are centered near $\alpha=1$, and how much weight they put on small $\alpha$ depends on the lower edge of the prior.

Our previous analysis without reheating found a larger value for Planck, $\log_{10}\alpha=1.32^{+0.69}_{-0.91}$~\cite{Carrion2026}. The main reason is the prior. There $r$ had a uniform prior, which places most of the prior volume at large $\alpha$, whereas here the prior is uniform in $\log_{10}\alpha$. If we reweight our no-reheating Planck chain to the old prior, with an approximate slow-roll change of variables, the median of $\log_{10}\alpha$ moves from $0.4$ to $0.9$. The rest of the difference comes from the different Planck likelihoods and the slow-roll treatment used there.

On the reheating side the picture is the same as at $\alpha=1$. P-ACT+DESI gives $\wrh=0.75^{+0.14}_{-0.20}$, $\ln R_{\rm rad}=6^{+3}_{-4}$ and $\log_{10}(\Trh/{\rm GeV})<11.0$, and P-ACT gives $\wrh=0.65^{+0.22}_{-0.30}$ and $\log_{10}(\Trh/{\rm GeV})<13.4$, all at 95\% for the temperature limits. As at $\alpha=1$, the temperature limits are prior-dominated.

\begin{figure*}[t]
\centering
\includegraphics[width=0.72\textwidth]{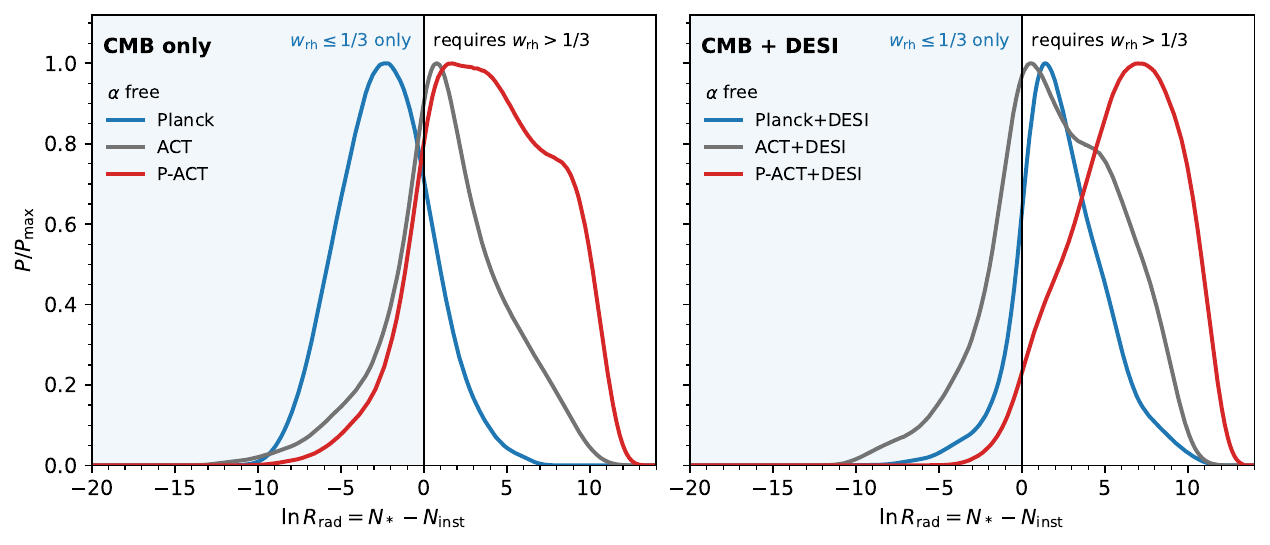}
\caption{As \cref{fig:lnR}, for free $\alpha$. Since $N_{\rm inst}$ now depends on $\alpha$, there is no one-to-one map to $\Nstar$ and the upper axis is omitted.}
\label{fig:lnR-free}
\end{figure*}

\begin{figure}[tbp]
\centering
\includegraphics[width=\columnwidth]{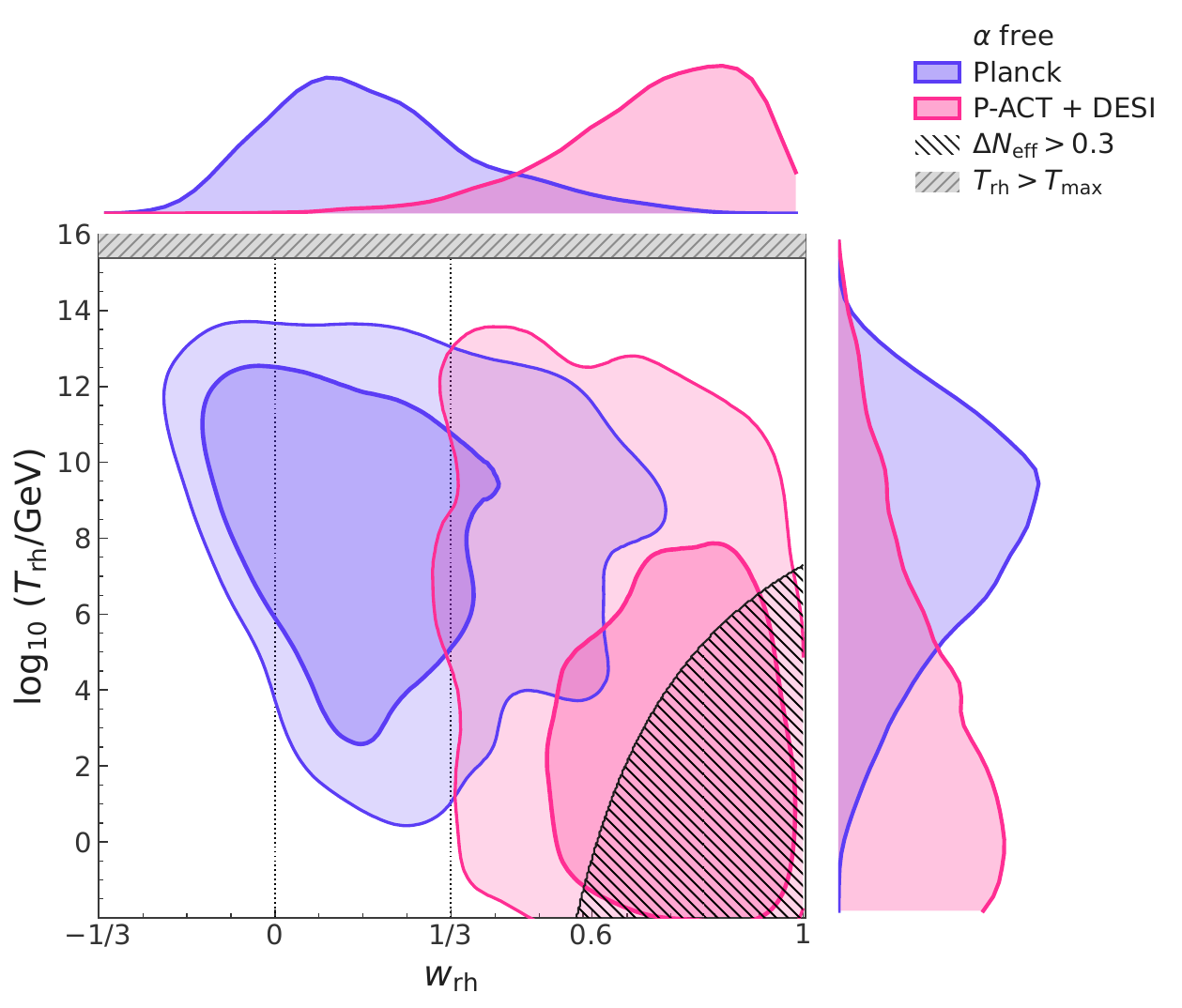}
\caption{As \cref{fig:plane}, for free $\alpha$. The band $\Trh>\Trhmax$ uses the median $\log_{10}(\Trhmax/{\rm GeV})=15.39$ of the P-ACT+DESI posterior, which varies by less than $0.1$~dex across it, and the gravitational-wave region is computed for $r=0.019$, the median of the same posterior. As in \cref{fig:plane}, the region is only indicative, here all the more because $r$ varies by an order of magnitude across the posterior.}
\label{fig:plane-free}
\end{figure}

\Cref{fig:lnR-free,fig:plane-free} repeat \cref{fig:lnR,fig:plane} with $\alpha$ free. All combinations except Planck alone still have positive medians of $\ln R_{\rm rad}$. P-ACT does so already without DESI, $\ln R_{\rm rad}=4\pm4$, and adding DESI moves it to $6^{+3}_{-4}$, so P-ACT+DESI still prefers a stiff phase. Adding DESI to Planck gives $\ln R_{\rm rad}=2.1^{+2.8}_{-2.1}$.

Planck alone goes the other way, to $\ln R_{\rm rad}=-2.3\pm2.8$ and $\wrh=0.16^{+0.22}_{-0.16}$. A larger $\alpha$ raises $\ns$ at fixed $\Nstar$, so with $\alpha$ free Planck reaches its preferred $\ns$ with a smaller $\Nstar$, that is, with $\ln R_{\rm rad}<0$, and matter-like reheating becomes allowed. It is also the only combination that bounds $\alpha$ from below, $\log_{10}\alpha>-0.22$ at 95\%, which even Planck+DESI does not. The gravitational-wave region is slightly larger than at $\alpha=1$ because the preferred $\alpha$ implies a larger $r$.

\begin{figure*}[t]
\centering
\includegraphics[width=0.65\textwidth]{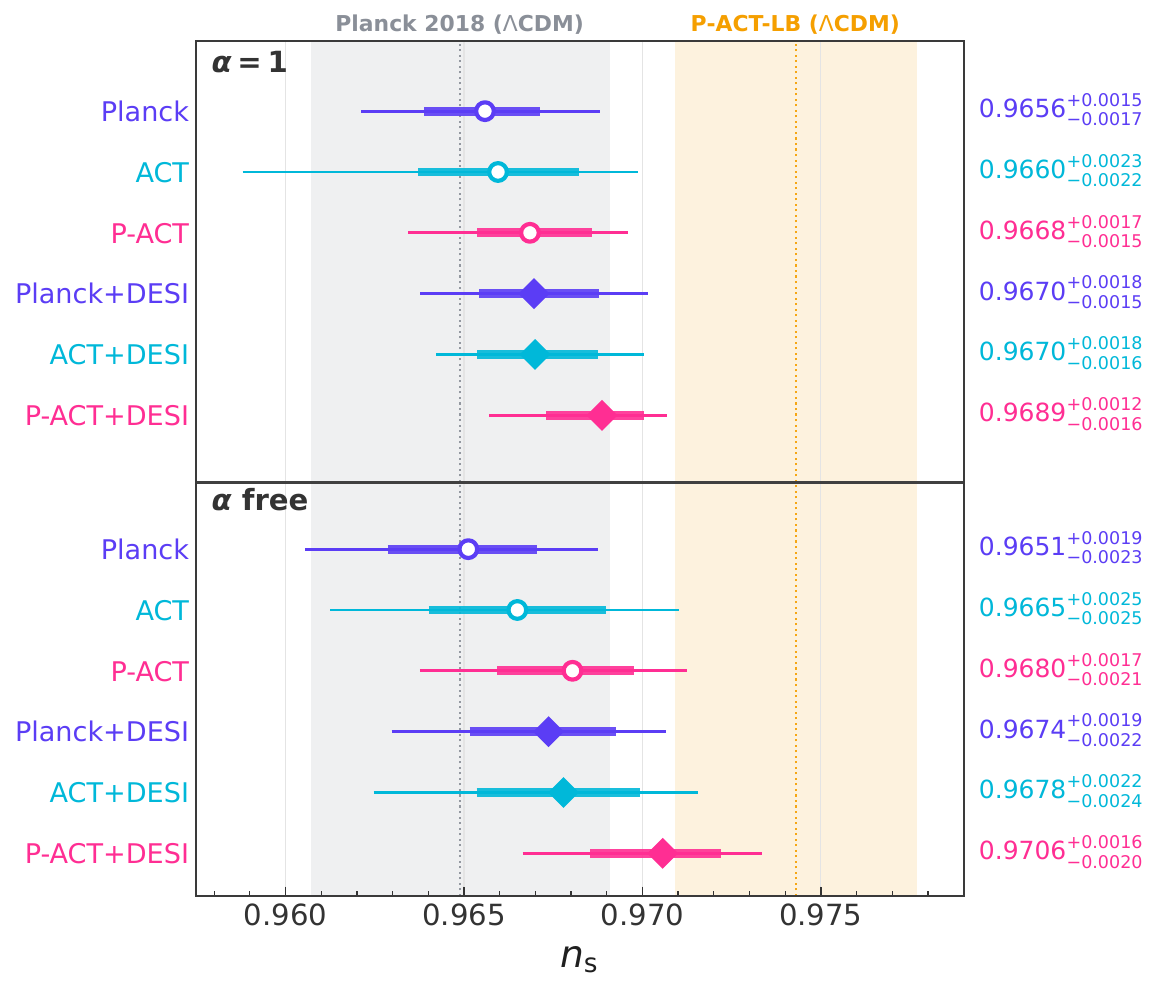}
\caption{Marginalized constraints on the scalar spectral index for the six data combinations, for the canonical model (upper panel) and with free $\alpha$ (lower panel). Open circles denote the CMB-only combinations and filled diamonds their DESI counterparts. The grey and orange vertical bands show the $\Lambda$CDM constraints from Planck 2018~\cite{Planck2018VI} and P-ACT-LB~\cite{Louis2025}. The latter includes CMB lensing and DESI DR1 BAO, and both differ from the likelihoods used here.}
\label{fig:ns-all}
\end{figure*}

\Cref{fig:ns-all} compares $\ns$ for all twelve runs. In both models, each combination with DESI gives a larger $\ns$ than the same combination without it. Freeing $\alpha$ raises $\ns$ appreciably only for P-ACT+DESI, while for the other combinations the change is at most $0.0012$.

\subsection{Reheating vs.\ no reheating}
\label{sec:impact_reheating}

\begin{figure*}[t]
\centering
\includegraphics[width=0.80\textwidth]{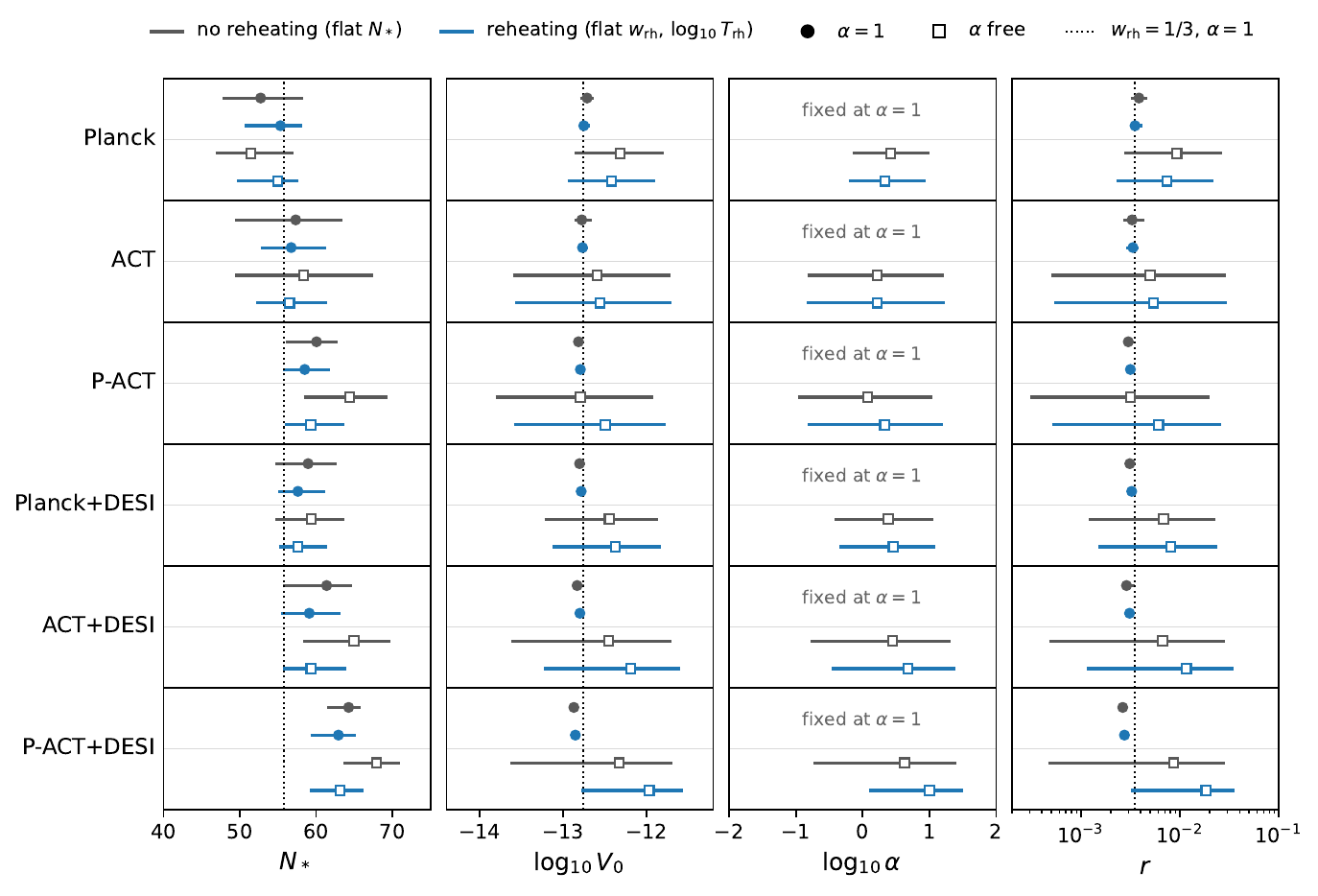}
\caption{Inflationary parameters and tensor-to-scalar ratio with and without reheating, for the same likelihood. Median and $68\%$ interval for each data combination. Grey: no reheating, with a uniform prior on $\Nstar$. Blue: reheating, with uniform priors on $\wrh\in[-1/3,1]$ and $\log_{10}\Trh$. In each panel the upper pair (filled circles) is $\alpha=1$ and the lower pair (open squares) is free $\alpha$. The dotted lines mark $\Nstar=N_{\rm inst}$ at $\alpha=1$ (instantaneous reheating, or equivalently $\wrh=1/3$).}
\label{fig:impact}
\end{figure*}

To isolate the effect of the reheating sector, we compare with an analysis without it, in which $\Nstar$ is a free parameter with a uniform prior on $[20,69]$ for $\alpha=1$ and on $[20,75]$ for free $\alpha$. The potential, the cosmological parameters, the likelihoods, the \texttt{CLASS} settings and the sampler are the same.  This comparison is not entirely straightforward and was performed with some subtleties; in particular, it relies on the reweighting procedure described below. Since, by \cref{eq:DeltaN}, at fixed $V_0$ and $\alpha$ the pair $(\wrh,\Trh)$ only selects a value of $\Nstar$, the two analyses share the same likelihood and differ only in the prior on $\Nstar$. A no-reheating chain therefore becomes the reheating posterior by reweighting each sample,
\begin{equation}
p_{\rm reh}(\theta,\Nstar\,|\,\mathcal D)\;\propto\;p_{\rm none}(\theta,\Nstar\,|\,\mathcal D)\,\frac{\pi_{\rm reh}(\Nstar\,|\,\alpha,V_0)}{\pi_{\rm none}(\Nstar)},
\label{eq:reweight}
\end{equation}
where $\pi_{\rm reh}$ is the prior on $\Nstar$ induced by the priors on $(\wrh,\Trh)$ and the conditions of \cref{tab:priors}.\footnote{We checked this on $287$ points drawn from the prior and the posterior: at the same $(V_0,\Nstar)$ and remaining parameters, the two models give the same $\ns$ to $10^{-5}$, and their $\ln\mathcal L$ differ by a median of $3\times10^{-4}$ and by less than $0.45$ within $\Delta\chi^2<30$ of the best fit, which is negligible compared with the variation of $\ln\mathcal L$ across the posterior.} We compute $\pi_{\rm reh}$ analytically. At fixed $(\alpha,V_0)$, \cref{eq:DeltaN} gives $\Nstar=N_{\rm inst}+f(\wrh)\ln(\Trh/\Trhmax)$ with $f(w)=(1-3w)/[3(1+w)]$, so the uniform prior on $(\wrh,\log_{10}\Trh)$ maps onto a closed-form density in $\Nstar-N_{\rm inst}$, which agrees with prior samples from \texttt{CLASS}. We fit $N_{\rm inst}$ and $\Trhmax$ as functions of $(\log_{10}V_0,\log_{10}\alpha)$. On the direct reheating chains the fit residuals are below $0.05$ $e$-folds and $10^{-3}$~dex, and shifting $N_{\rm inst}$ by $0.06$ $e$-folds changes the reweighted medians and widths by less than $0.02\sigma$. The reheating columns of \cref{tab:reh_vs_noreh} are obtained in this way, and \cref{fig:impact} compares the two analyses for the six data combinations.

\textit{Direct sampling and reweighting.} We therefore have two ways of obtaining the reheating posteriors: sampling the reheating model directly, as in \cref{sec:canonical,sec:alpha-general}, or reweighting a no-reheating chain with \cref{eq:reweight}. The direct runs, however, give narrower posteriors. The width of $\ns$ is smaller for Planck and for ACT+DESI at $\alpha=1$, appreciably smaller for Planck, ACT and P-ACT with free $\alpha$, and only marginally smaller in the other cases. The relevant distinction lies with the direct runs. In the $(\wrh,\Trh)$ plane the posterior is a long, thin band along a line of constant $\ln R_{\rm rad}$, which is hard for the sampler to fill, and the part it misses is the low-$\Nstar$ end of the band. More particles fill this part in gradually, but not completely, as shown in \cref{app:sampler-validation}. The reweighting does not have this difficulty, because there $\Nstar$ is sampled directly, and a no-reheating chain with a uniform prior on $\ns$ instead of $\Nstar$ gives the same widths. We still quote the direct runs in \cref{tab:alpha1-results,tab:alpha-free-results}, because they constrain $\wrh$ and $\Trh$ themselves and include the restrictive prior. 

\begin{table*}[t]
\centering
\footnotesize
\setlength{\tabcolsep}{2.5pt}
\renewcommand{\arraystretch}{1.45}
\begin{tabular}{lcccccccc}
\toprule
& \multicolumn{2}{c}{$\Nstar$} & \multicolumn{2}{c}{$\log_{10}V_0$} & \multicolumn{2}{c}{$\log_{10}\alpha$} & \multicolumn{2}{c}{$10^3\,r$} \\
\cmidrule(lr){2-3}\cmidrule(lr){4-5}\cmidrule(lr){6-7}\cmidrule(lr){8-9}
Data & none & reh. & none & reh. & none & reh. & none & reh. \\
\midrule
\multicolumn{9}{l}{\textit{$\alpha=1$}} \\
Planck & $53^{+6}_{-5}$ & $55.3^{+2.9}_{-4.7}$ & $-12.71\pm0.08$ & $-12.75^{+0.07}_{-0.04}$ & -- & -- & $3.8^{+0.8}_{-0.7}$ & $3.5^{+0.6}_{-0.3}$ \\
ACT & $57^{+6}_{-8}$ & $57^{+5}_{-4}$ & $-12.78^{+0.13}_{-0.09}$ & $-12.77^{+0.06}_{-0.07}$ & -- & -- & $3.3^{+1.1}_{-0.6}$ & $3.3\pm0.5$ \\
P-ACT & $60.1^{+2.8}_{-4.0}$ & $58.5^{+3.4}_{-2.9}$ & $-12.82^{+0.06}_{-0.04}$ & $-12.79^{+0.04}_{-0.05}$ & -- & -- & $3.00^{+0.42}_{-0.25}$ & $3.2\pm0.3$ \\
Planck+DESI & $59\pm4$ & $57.6^{+3.6}_{-2.5}$ & $-12.80^{+0.06}_{-0.05}$ & $-12.78^{+0.04}_{-0.05}$ & -- & -- & $3.1^{+0.5}_{-0.3}$ & $3.25^{+0.29}_{-0.36}$ \\
ACT+DESI & $61^{+3}_{-6}$ & $59\pm4$ & $-12.83^{+0.09}_{-0.05}$ & $-12.80\pm0.06$ & -- & -- & $2.88^{+0.59}_{-0.28}$ & $3.1\pm0.4$ \\
P-ACT+DESI & $64.3^{+1.6}_{-2.8}$ & $62.9^{+2.3}_{-3.7}$ & $-12.872^{+0.038}_{-0.021}$ & $-12.85^{+0.05}_{-0.03}$ & -- & -- & $2.64^{+0.23}_{-0.12}$ & $2.74^{+0.33}_{-0.18}$ \\
\addlinespace[2pt]
\multicolumn{9}{l}{\textit{$\alpha$ free}} \\
Planck & $51^{+6}_{-5}$ & $54.9^{+2.8}_{-5.3}$ & $-12.3\pm0.5$ & $-12.4\pm0.5$ & $0.4\pm0.6$ & $0.3^{+0.6}_{-0.5}$ & $9^{+17}_{-7}$ & $7^{+14}_{-5}$ \\
ACT & $58\pm9$ & $57^{+5}_{-4}$ & $-12.6^{+0.9}_{-1.0}$ & $-12.6^{+0.9}_{-1.0}$ & $0.2^{+1.0}_{-1.1}$ & $0.2^{+1.0}_{-1.1}$ & $5^{+24}_{-4}$ & $5^{+24}_{-5}$ \\
P-ACT & $64^{+5}_{-6}$ & $59^{+4}_{-3}$ & $-12.8^{+0.9}_{-1.0}$ & $-12.5^{+0.7}_{-1.1}$ & $0.1\pm1.0$ & $0.3^{+0.9}_{-1.1}$ & $3.2^{+16.9}_{-2.8}$ & $6^{+20}_{-6}$ \\
Planck+DESI & $59^{+4}_{-5}$ & $57.6^{+3.8}_{-2.5}$ & $-12.4^{+0.6}_{-0.8}$ & $-12.4^{+0.5}_{-0.8}$ & $0.4^{+0.7}_{-0.8}$ & $0.5^{+0.6}_{-0.8}$ & $7^{+16}_{-6}$ & $8^{+16}_{-7}$ \\
ACT+DESI & $65^{+5}_{-7}$ & $59^{+5}_{-4}$ & $-12.5^{+0.8}_{-1.2}$ & $-12.2^{+0.6}_{-1.0}$ & $0.4^{+0.9}_{-1.2}$ & $0.7^{+0.7}_{-1.1}$ & $7^{+22}_{-6}$ & $12^{+23}_{-10}$ \\
P-ACT+DESI & $68^{+3}_{-4}$ & $63^{+3}_{-4}$ & $-12.3^{+0.6}_{-1.3}$ & $-12.0^{+0.4}_{-0.8}$ & $0.6^{+0.8}_{-1.4}$ & $1.0^{+0.5}_{-0.9}$ & $9^{+20}_{-8}$ & $18^{+18}_{-15}$ \\
\bottomrule
\end{tabular}
\caption{Inflationary parameters and $r$ without reheating (``none'', uniform prior on $\Nstar$) and with reheating (``reh.'', uniform priors on $\wrh\in[-1/3,1]$ and $\log_{10}\Trh$), for the same likelihood. Medians and $68\%$ equal-tail intervals. The reheating columns are obtained from the no-reheating chains with \cref{eq:reweight}.}
\label{tab:reh_vs_noreh}
\end{table*}

The prior that the reheating sector places on $\Nstar$ differs from a uniform one in two ways. First, it has hard limits. No history with $\wrh\le1$ and $\Trh\ge\TBBN$ can move $\Nstar$ outside $-L\le\ln R_{\rm rad}\le L/3$, with $L=\ln(\Trhmax/\TBBN)$, which is about $[-40,13.3]$ at $\alpha=1$. Second, it is strongly peaked at $\Nstar=N_{\rm inst}$. Every point on the line $\wrh=1/3$, whatever $\Trh$, and every point with $\Trh=\Trhmax$, whatever $\wrh$, gives exactly $\ln R_{\rm rad}=0$ according to \cref{eq:DeltaN}, and the points near these two lines give values close to zero. As a result, $66\%$ of the prior lies in $50<\Nstar<62$, compared with $25\%$ for a uniform prior on $[20,69]$. The limits remove values of $\Nstar$ that no reheating history can produce, and the peak pulls the posterior toward instantaneous reheating.

At $\alpha=1$ the peak dominates. Reheating narrows the posteriors of $\Nstar$ and $r$ for the first several combinations and moves them toward $N_{\rm inst}$, by up to a modest number of $e$-folds. P-ACT+DESI is the exception. Its likelihood lies well above $N_{\rm inst}$, the peak pulls a tail toward it, and its posterior becomes wider. If we replace the reheating prior by one that is uniform in $\ln R_{\rm rad}$ over the same physical range, which keeps the limits but removes the peak, all widths return to their no-reheating values. Thus, the reheating sector adds a preference for $\Nstar\simeq N_{\rm inst}$, not new information from the data.

On the other hand, with $\alpha$ free the limits matter more. For the ACT-based combinations, a substantial fraction of the no-reheating posterior lies beyond the largest $\Nstar$ that any reheating history can reach. Reheating removes this part, which lowers $\Nstar$ by several $e$-folds for P-ACT, ACT+DESI and P-ACT+DESI. These combinations then reach their large $\ns$ through a larger $\alpha$, and hence a larger $r$. For P-ACT+DESI the median of $r$ roughly doubles.

\subsection{Is a stiff phase required?}
\label{subsec:restrictive}

A direct test is to forbid the stiff branch. We repeat the free-$\alpha$ analysis with $-1/3<\wrh\le1/3$ and change nothing else. A similar test appears in~\cite{Ballardini2025}, whose restrictive prior also requires $\rho_{\rm rh}^{1/4}>1$~TeV, while we keep our BBN floor.

Without a stiff phase there is a maximum value of $\ns$. For $\wrh\le1/3$, \cref{eq:DeltaN} gives $\ln R_{\rm rad}\le0$, so $\Nstar$ cannot exceed $N_{\rm inst}(\alpha)$ and $\ns$ cannot exceed the value it takes for instantaneous reheating. Computed with \texttt{CLASS} at fixed $A_s$, this maximum is $0.9653$ at $\alpha=1$ and grows slowly with $\alpha$, from $0.9638$ at $\alpha=0.1$ to $0.9678$ at $\alpha=10$ and $0.9685$ at $\alpha=50$. Beyond that it hardly changes, and $r$ is already excluded: for instantaneous reheating, $r$ reaches the BICEP/Keck bound $r=0.036$~\cite{BK18} at $\alpha\simeq24$. A stiff phase raises the maximum. For the most extreme history allowed by our priors, $\wrh=1$ with $\Trh=10$~MeV, it becomes $0.9718$ at $\alpha=1$, and $0.9707$ to $0.9745$ between $\alpha=0.1$ and $50$.

\begin{figure*}[t]
\centering
\includegraphics[width=0.65\textwidth]{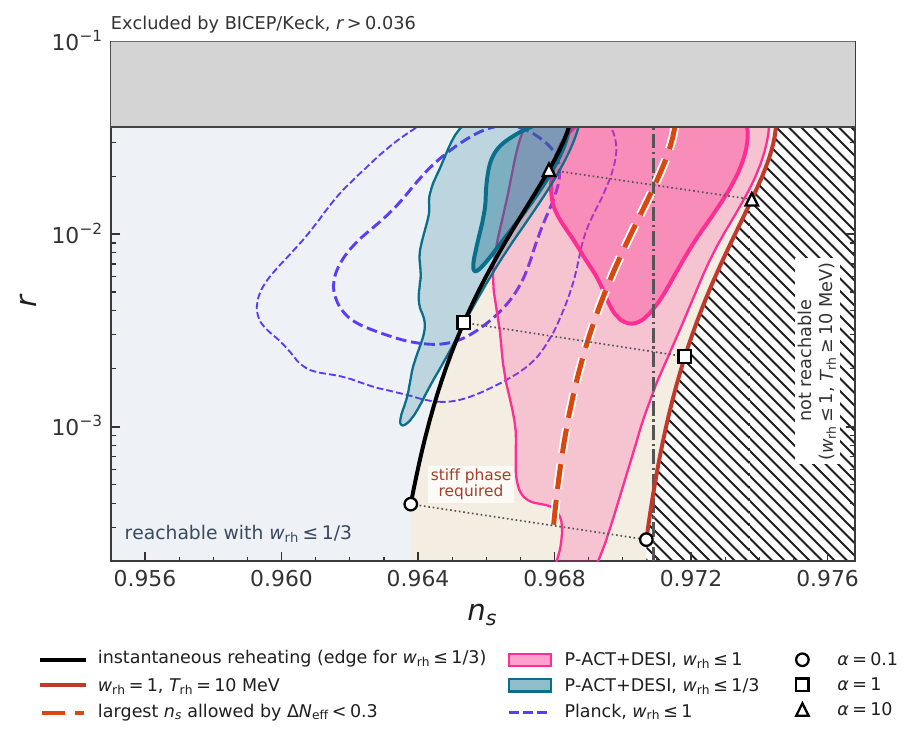}
\caption{Reach of the reheating histories in the $(\ns,r)$ plane for $0.1\le\alpha\le50$, computed with \texttt{CLASS} at fixed $A_s$. Black: instantaneous reheating, the edge of the region reachable with $\wrh\le1/3$ (light blue, to its left). Red: the most extreme history allowed by our priors, $\wrh=1$ with $\Trh=10$~MeV. The band between the two curves requires a stiff phase, and the hatched region on the right cannot be reached. Dashed orange: approximate largest $\ns$ allowed by $\Delta N_{\rm eff}<0.3$. Markers joined by dotted lines show the two edges for $\alpha=0.1$, $1$ and $10$. Below $\alpha=0.1$, where $\ns$ barely changes with $\alpha$, the regions are continued as vertical bands. Contours ($68\%$ and $95\%$): P-ACT+DESI with free $\alpha$ and the permissive prior (magenta) or the restrictive prior $\wrh\le1/3$ (teal), and Planck with the permissive prior (violet dashed). The grey band at the top is excluded by BICEP/Keck, $r>0.036$, and the vertical dash-dotted line marks the lower end of the $1\sigma$ P-ACT-LB $\Lambda$CDM interval, $\ns=0.9709$.}
\label{fig:ceiling}
\end{figure*}

\Cref{fig:ceiling} shows what this means in the $(\ns,r)$ plane. Everything to the left of the instantaneous-reheating curve can be reached with $\wrh\le1/3$, and the band between it and the curve for $\wrh=1$ requires a stiff phase. Planck lies mostly in the first region. Most of the P-ACT+DESI posterior lies in the second, and its median, $\ns=0.9706$, is above the largest value that any $\alpha\le50$ can reach without a stiff phase. Part of this posterior also lies above the BICEP/Keck bound: P-ACT+DESI allows $r<0.043$ at $95\%$, above $r<0.036$, because none of our combinations includes $B$-mode data. It can only be reached with a stiff phase, for $\alpha\gtrsim0.2$, and its central value only for the most extreme history at $\alpha\simeq22$, where $r$ is already at the BICEP/Keck bound.

\begin{figure*}[t]
\centering
\includegraphics[width=0.65\textwidth]{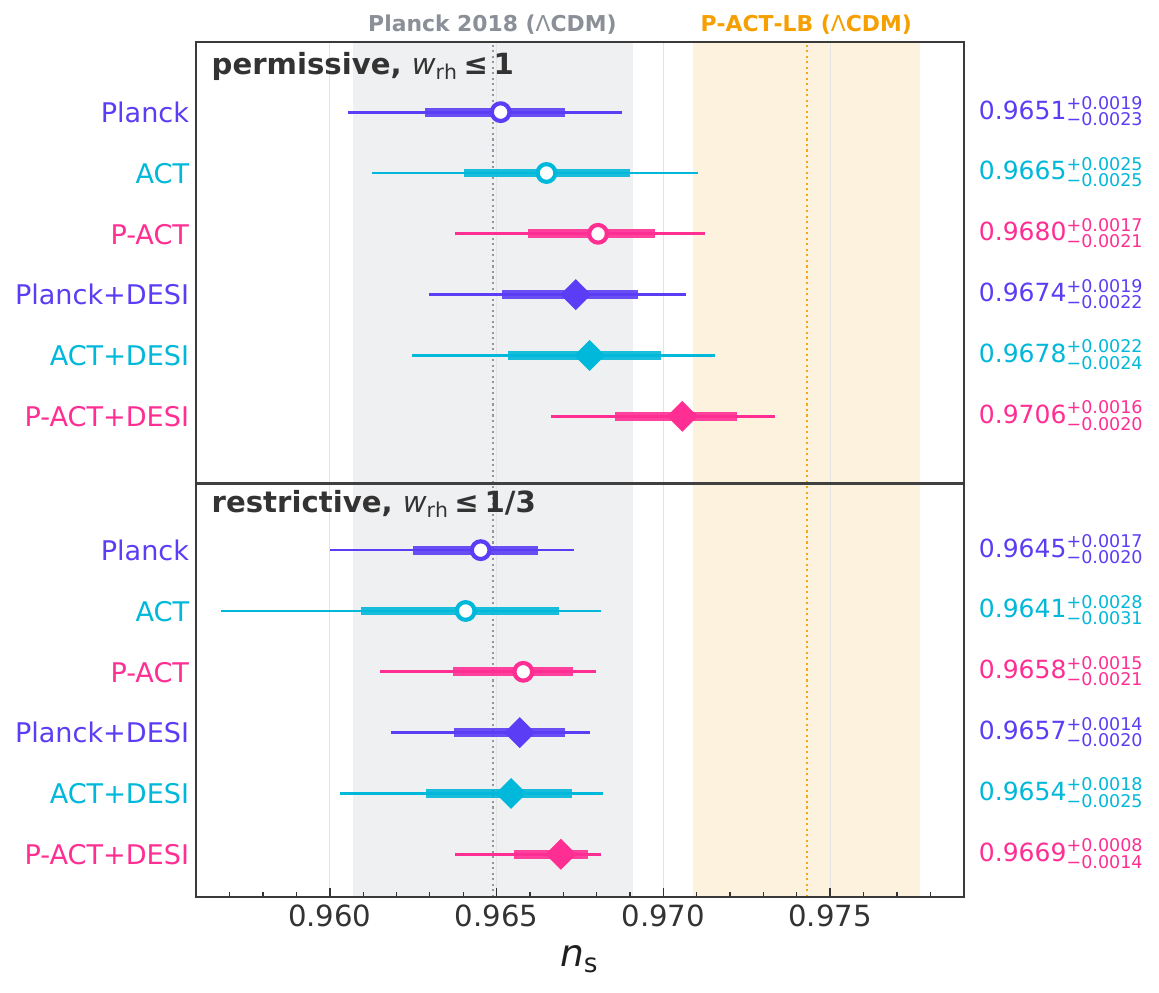}
\caption{As \cref{fig:ns-all}, for free $\alpha$ with the permissive prior, $-1/3<\wrh\le1$ (upper panel), and the restrictive prior, $-1/3<\wrh\le1/3$ (lower panel).}
\label{fig:ns-restr}
\end{figure*}

Forbidding the stiff phase confirms this picture (\cref{tab:restrictive}, \cref{fig:ceiling,fig:ns-restr}). For Planck the restriction changes little, $\ns=0.9645^{+0.0017}_{-0.0020}$ and $\Nstar=52.5^{+2.1}_{-2.8}$, compared with $0.9651^{+0.0019}_{-0.0023}$ and $53.8\pm2.8$. For P-ACT+DESI the posterior is pushed against the maximum. $\Nstar$ piles up just below $N_{\rm inst}$, at $55.4^{+0.7}_{-1.4}$ instead of $63^{+3}_{-4}$, and $\ns$ falls from $0.9706^{+0.0016}_{-0.0020}$ to $0.9669^{+0.0008}_{-0.0014}$. To get as close as possible to the maximum with $\wrh\le1/3$, reheating must be almost instantaneous, so the posterior now prefers high reheating temperatures, $\log_{10}(\Trh/{\rm GeV})=11^{+3}_{-6}$, instead of the upper bound found with the permissive prior. In the other five combinations $\alpha$ moves up, because a larger $\alpha$ raises both $\ns$ at fixed $\Nstar$ and the maximum itself. With the restrictive prior all six combinations stay below the maximum and the DESI combinations lose most of their upward shift.

\begin{table*}[t]
\centering
\footnotesize
\setlength{\tabcolsep}{6pt}
\renewcommand{\arraystretch}{1.4}
\begin{tabular}{lccccc}
\toprule
 & $\ns$ & $\Nstar$ & $\wrh$ & $\log_{10}(\Trh/{\rm GeV})$ & $\log_{10}\alpha$ \\
\midrule
\multicolumn{6}{l}{\textit{Permissive prior: $-1/3<\wrh\le1$}} \\
Planck & $0.9651^{+0.0019}_{-0.0023}$ & $53.8\pm2.8$ & $0.16^{+0.22}_{-0.16}$ & $8.4^{+2.6}_{-3.3}$ & $0.4^{+0.5}_{-0.4}$ \\
ACT & $0.9665\pm0.0025$ & $57^{+4}_{-3}$ & $0.47^{+0.28}_{-0.26}$ & $7\pm5$ & $0.1^{+0.7}_{-0.8}$ \\
P-ACT & $0.9680^{+0.0017}_{-0.0021}$ & $60\pm4$ & $0.65^{+0.22}_{-0.30}$ & $<13.4$ & $0.1^{+0.8}_{-1.1}$ \\
Planck+DESI & $0.9674^{+0.0019}_{-0.0022}$ & $58.0^{+2.8}_{-2.2}$ & $0.57^{+0.22}_{-0.23}$ & $8^{+4}_{-5}$ & $0.2\pm0.7$ \\
ACT+DESI & $0.9678^{+0.0022}_{-0.0024}$ & $58^{+4}_{-3}$ & $0.49^{+0.24}_{-0.28}$ & $<13.1$ & $0.6^{+0.5}_{-0.8}$ \\
P-ACT+DESI & $0.9706^{+0.0016}_{-0.0020}$ & $63^{+3}_{-4}$ & $0.75^{+0.14}_{-0.20}$ & $<11.0$ & $1.0^{+0.5}_{-0.8}$ \\
\midrule
\multicolumn{6}{l}{\textit{Restrictive prior: $-1/3<\wrh\le1/3$}} \\
Planck & $0.9645^{+0.0017}_{-0.0020}$ & $52.5^{+2.1}_{-2.8}$ & $0.03\pm0.11$ & $9.9^{+2.5}_{-3.2}$ & $0.7^{+0.5}_{-0.6}$ \\
ACT & $0.9641^{+0.0028}_{-0.0031}$ & $52.8^{+2.4}_{-4.1}$ & $0.10^{+0.14}_{-0.19}$ & $8^{+4}_{-5}$ & $0.5^{+0.8}_{-1.0}$ \\
P-ACT & $0.9658^{+0.0015}_{-0.0021}$ & $54.2^{+1.3}_{-2.2}$ & $0.15^{+0.11}_{-0.16}$ & $9^{+4}_{-5}$ & $0.8^{+0.4}_{-0.7}$ \\
Planck+DESI & $0.9657^{+0.0014}_{-0.0020}$ & $53.8^{+1.3}_{-2.0}$ & $0.13^{+0.11}_{-0.15}$ & $8^{+3}_{-4}$ & $0.8^{+0.5}_{-0.7}$ \\
ACT+DESI & $0.9654^{+0.0018}_{-0.0025}$ & $54.2^{+1.5}_{-2.8}$ & $0.15^{+0.12}_{-0.18}$ & $10^{+4}_{-6}$ & $0.7^{+0.6}_{-1.0}$ \\
P-ACT+DESI & $0.9669^{+0.0008}_{-0.0014}$ & $55.4^{+0.7}_{-1.4}$ & $0.19^{+0.10}_{-0.16}$ & $11^{+3}_{-6}$ & $0.94^{+0.27}_{-0.46}$ \\
\bottomrule
\end{tabular}
\caption{Free $\alpha$: permissive ($-1/3<\wrh\le1$) versus restrictive ($-1/3<\wrh\le1/3$) reheating prior. Medians and 68\% equal-tail intervals, rounded as in \cref{tab:alpha1-results}. Upper limits are one-sided 95\% bounds.}
\label{tab:restrictive}
\end{table*}

The stiff phase, however, is where the gravitational-wave bound becomes relevant. Imposing our approximation on $\Delta N_{\rm eff}<0.3$ removes $53\%$ of the P-ACT+DESI posterior with free $\alpha$ ($49\%$ at $\alpha=1$). The removed part is the corner of large $\wrh$ and low $\Trh$ that the data prefer most (\cref{fig:plane,fig:plane-free}). In \cref{fig:ceiling} the bound appears as the dashed orange edge: with $\Delta N_{\rm eff}<0.3$, $\ns$ cannot exceed $0.9690$ at $\alpha=1$ or $0.9715$ for $\alpha\le50$, so only about half of the stiff band survives. Dedicated studies point the same way: the $1\sigma$ region of stiff reheating for Starobinsky inflation is already excluded by BBN~\cite{Dorsch2026}, and a kination era driven by a scalar field cannot last more than about $10$ $e$-folds~\cite{Eroncel2025}, whereas P-ACT+DESI prefers $\Nrh=22^{+6}_{-8}$. The CMB data thus favour a stiff phase, although this is not in agreement with the gravitational-wave bound. Until both are combined in the likelihood, the indication for $\wrh>1/3$ should not be read as a detection; the restrictive prior then provides a conservative reference.

\section{Illustrative perturbative inflaton decay scenarios}
\label{sec:microphysics}

The phenomenological reheating description adopted above can be connected to the microphysics of inflaton decay under additional assumptions. Perturbative reheating has been studied in Starobinsky-like and $\alpha$-attractor models, relating the inflaton decay rate to radiation production, the onset of radiation domination and the corresponding inflationary $e$-fold number~\cite{Ellis2015,Ellis2025,GermanPerturbative2025}. For the quadratic minimum of the potential considered here, coherent inflaton oscillations give a predominantly matter-like expansion. The effective equation of state averaged over the whole reheating epoch, \cref{eq:weff}, need not vanish exactly, however, because the produced radiation gradually contributes to the total energy density.

In this section we use representative perturbative decay channels to map the reheating temperatures inferred from our phenomenological analysis into indicative decay rates and couplings. These estimates are conditional on radiation domination being reached through decay, rather than a complete microscopic realization of every reheating history sampled. In particular, the stiff histories preferred by some data combinations cannot be generated by perturbative decay of a coherently oscillating inflaton in the quadratic minimum alone.

\subsection{Decay rates and couplings}

If radiation takes over when $H\simeq\Gamma_\phi$, the reheating temperature is
\begin{equation}
\Trh\simeq\left(\frac{90}{\pi^2\grh}\right)^{1/4}\sqrt{\Gamma_\phi\mpl},
\label{eq:TGamma}
\end{equation}
with $\Gamma_\phi$ the total decay rate and $\grh=106.75$. This is a leading-order estimate that identifies the transition to radiation domination with $H\sim\Gamma_\phi$ and assumes sufficiently rapid thermalization. More detailed treatments follow the continuous transfer of energy to radiation and can give numerical corrections to this relation, depending in part on how the end of reheating is defined. We retain \cref{eq:TGamma} as a representative mapping, rather than a unique reconstruction of the decay dynamics, since these normalization uncertainties are smaller than the broad inferred ranges of $\Trh$.

The inflaton mass is the curvature of the potential at its minimum,
\begin{equation}
m_\phi^2=V''(0)=\frac{4}{3\alpha}\frac{V_0}{\mpl^2}.
\label{eq:mphi}
\end{equation}
We consider three channels,
\begin{align}
\Gamma_\phi^{\rm grav}&\simeq\frac{m_\phi^3}{192\pi\mpl^2},\label{eq:Ggrav}\\
\Gamma_{\phi\rightarrow\chi\chi}&\simeq\frac{g^2}{8\pi m_\phi},\label{eq:Gscalar}\\
\Gamma_{\phi\rightarrow\bar\psi\psi}&\simeq\frac{y^2m_\phi}{8\pi}.\label{eq:Gyuk}
\end{align}
The first is the decay of the $R^2$ scalaron into one light real scalar minimally coupled in the Jordan frame~\cite{GorbunovPanin2011}. At free $\alpha$ we keep it as a benchmark, since the potential alone does not fix the matter couplings. The other two correspond to a trilinear coupling $g\,\phi\chi^2$ and a Yukawa coupling $y\,\phi\bar\psi\psi$~\cite{Allahverdi2010,Garcia2020}.

For every sample in the chains we compute $m_\phi$ from $(V_0,\alpha)$ and $\Gamma_\phi$ from $\Trh$. If a single channel supplies the whole rate, the couplings are
\begin{equation}
\begin{gathered}
y=\sqrt{\frac{8\pi\Gamma_\phi}{m_\phi}},
\qquad
g=\sqrt{8\pi\Gamma_\phi\,m_\phi},
\\
\frac{\Gamma_\phi}{m_\phi}=\frac{y^2}{8\pi}=\frac{1}{8\pi}\left(\frac{g}{m_\phi}\right)^2.
\end{gathered}
\label{eq:couplings}
\end{equation}
The gravitational rate needs no free coupling: it is fixed by $m_\phi$, and so is the corresponding temperature $\Trh^{\rm grav}$. For a given $\wrh$ this fixes $\Nrh=4\ln(\Trhmax/\Trh)/[3(1+\wrh)]$, and hence $\Nstar$ and $\ns$, which we read off a grid of \texttt{CLASS} runs (interpolated in $\alpha$ when $\alpha$ is free).

The data fix the inflaton mass well. For Planck and P-ACT+DESI, with $\alpha=1$ or free, $\log_{10}(m_\phi/{\rm GeV})$ lies between $13.37$ and $13.47$ with an uncertainty of at most $0.05$, so $m_\phi\simeq3\times10^{13}$~GeV. The gravitational channel then gives $\log_{10}(\Gamma_\phi^{\rm grav}/m_\phi)=-12.66\pm0.06$ at $\alpha=1$ and $-12.71^{+0.12}_{-0.16}$ with $\alpha$ free, which corresponds to $\Trh^{\rm grav}\simeq2\times10^{9}$~GeV, or to $y\simeq10^{-5.6}$ and $g\simeq10^{7.8}$~GeV if expressed through the other two channels.

\begin{figure*}[t]
\centering
\includegraphics[width=0.95\textwidth]{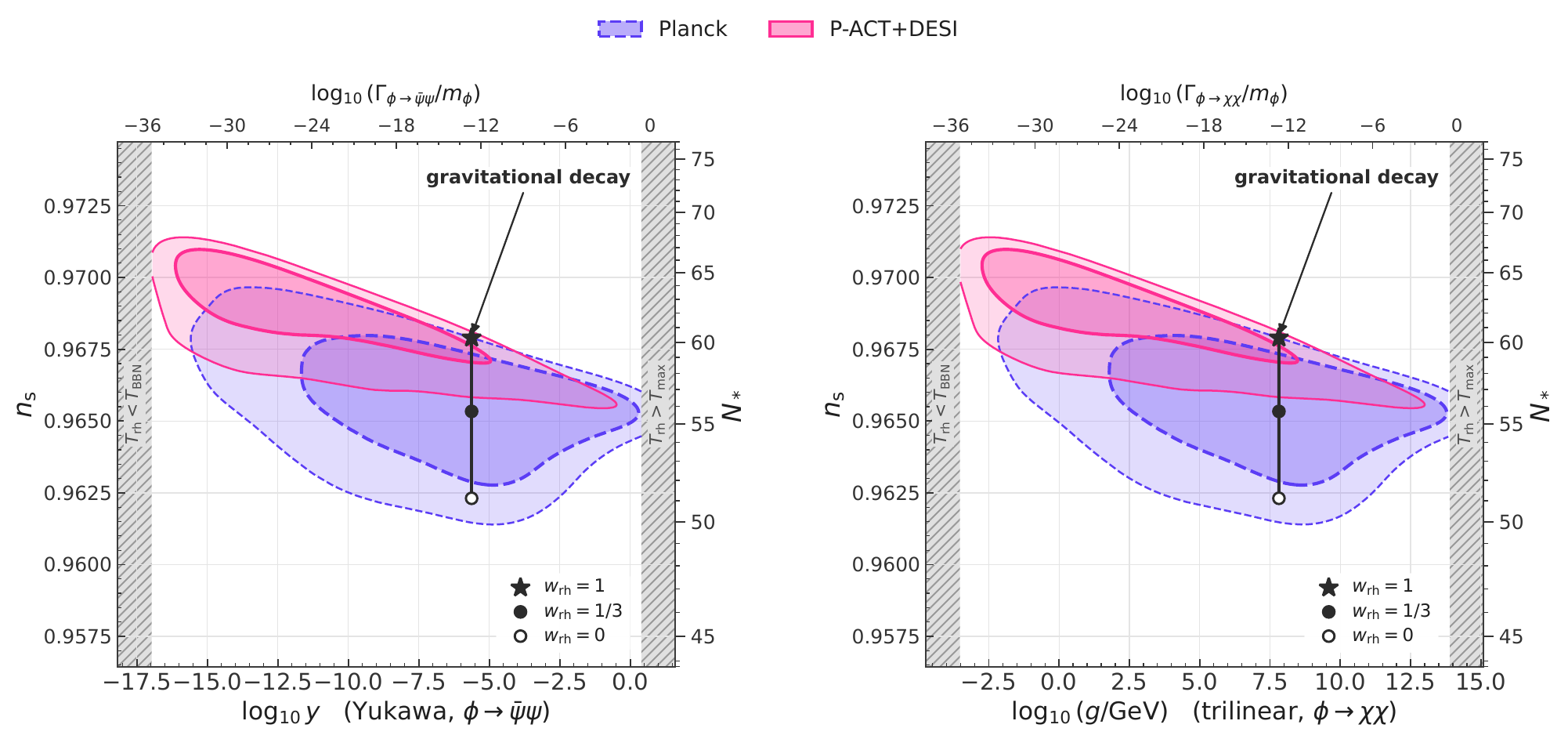}
\caption{Reconstructed inflaton couplings for $\alpha=1$: $68\%$ and $95\%$ contours of $(\log_{10}y,\,\ns)$ for Yukawa decay $\phi\to\bar\psi\psi$ (left) and $(\log_{10}(g/{\rm GeV}),\,\ns)$ for trilinear decay $\phi\to\chi\chi$ (right), for Planck (dashed) and P-ACT+DESI (solid). The couplings are obtained sample by sample from $\Trh$ through $H_{\rm rh}\simeq\Gamma_\phi$ and from $m_\phi=(4V_0/3\alpha)^{1/2}/\mpl$ of each sample. The right axes show $\Nstar$, in one-to-one correspondence with $\ns$ at $\alpha=1$, and the upper axes show $\Gamma_\phi/m_\phi$. Hatched bands: $\Trh<\TBBN$ and $\Trh>\Trhmax$. The black segment is the gravitational decay, with the $1\sigma$ width of the coupling reconstructed from $m_\phi$. Its markers give $\ns$ at the gravitational $\Trh$ for $\wrh=1$ ($\star$), $1/3$ ($\bullet$) and $0$ ($\circ$).}
\label{fig:couplings_alpha1}
\end{figure*}

\begin{figure*}[t]
\centering
\includegraphics[width=0.95\textwidth]{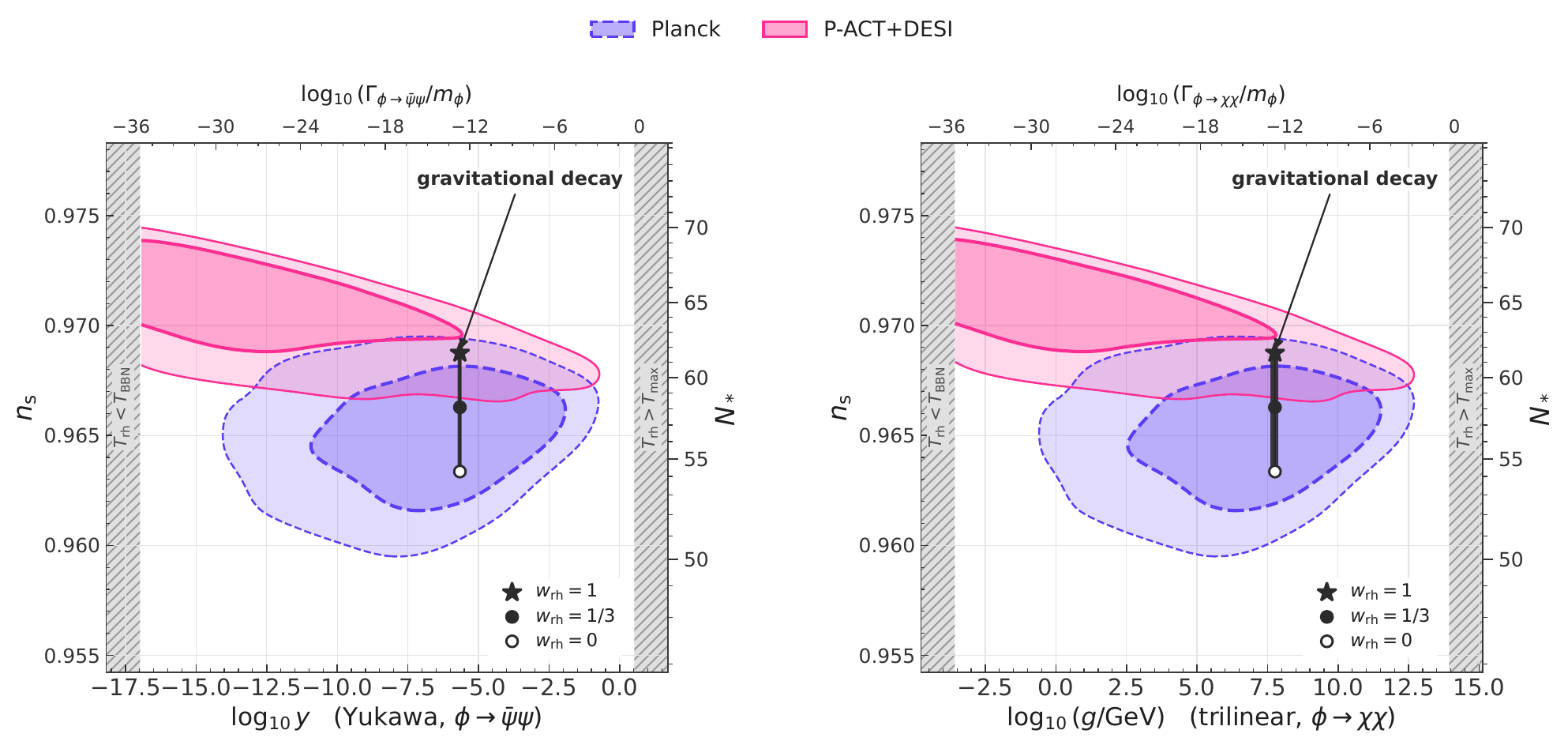}
\caption{As \cref{fig:couplings_alpha1}, for $\alpha$ free. With $\alpha$ free the same $\ns$ can correspond to different values of $\Nstar$, so the right axes give only the average $\Nstar$ of the chain samples at each $\ns$, with a spread of about $\pm1.6$ $e$-folds. The gravitational band is wider in the trilinear panel because $g_{\rm grav}\propto m_\phi^2$, whereas $y_{\rm grav}\propto m_\phi$.}
\label{fig:couplings_alphafree}
\end{figure*}

\Cref{fig:couplings_alpha1,fig:couplings_alphafree} show the reconstructed couplings against $\ns$ for Planck and P-ACT+DESI, and \cref{tab:couplings} lists them. The Yukawa and trilinear panels look alike because $\Gamma_\phi/m_\phi$ is the same function of $y$ and of $g/m_\phi$, and $m_\phi$ is essentially fixed. Their width is inherited from the prior-dominated posterior of $\Trh$. The P-ACT+DESI couplings come out some six orders of magnitude below the Planck ones, but the intervals are wide and overlap, and the shift is just the move toward low $\Trh$ along the stiff branch. We read these ranges as couplings compatible with the data, not as measurements.

\begin{table*}[t]
\centering
\footnotesize
\renewcommand{\arraystretch}{1.3}
\begin{tabular}{lcccc}
\toprule
 & \multicolumn{2}{c}{$\alpha=1$} & \multicolumn{2}{c}{$\alpha$ free} \\
\cmidrule(lr){2-3}\cmidrule(lr){4-5}
 & $\log_{10}y$ & $\log_{10}(g/{\rm GeV})$ & $\log_{10}y$ & $\log_{10}(g/{\rm GeV})$ \\
\midrule
Planck        & $-6^{+3}_{-5}$   & $7^{+3}_{-5}$   & $-6.6^{+2.6}_{-3.3}$  & $6.9^{+2.6}_{-3.3}$ \\
P-ACT+DESI    & $-12.0^{+4.5}_{-2.7}$  & $1.4^{+4.5}_{-2.7}$   & $-12^{+5}_{-3}$ & $1^{+5}_{-3}$ \\
Gravitational & $-5.63\pm0.03$   & $7.82^{+0.05}_{-0.06}$   & $-5.66^{+0.06}_{-0.08}$ & $7.76^{+0.12}_{-0.16}$ \\
\bottomrule
\end{tabular}
\caption{Reconstructed couplings (median and $68\%$ interval) for Yukawa and trilinear decay, and the values corresponding to gravitational decay. The Yukawa and trilinear ranges inherit the prior-dominated posterior of $\Trh$, so they are compatible couplings rather than measurements. The gravitational values depend only on $m_\phi$ and are computed from the same reheating chains, Planck and P-ACT+DESI together, which differ by only $0.04$ in $\log_{10}y$.}
\label{tab:couplings}
\end{table*}

The gravitational channel is the most interesting case, because it has no free coupling. It fixes $\Trh^{\rm grav}\simeq2\times10^9$~GeV, and for each $\wrh$ this fixes $\Nstar$ and $\ns$ (markers in \cref{fig:couplings_alpha1,fig:couplings_alphafree}). For matter-like reheating, $\wrh=0$, which is what perturbative decay of the oscillating inflaton gives, the prediction is $\ns=0.9623$ ($0.9634$ with $\alpha$ free). Planck agrees with it. This point lies inside the Planck $95\%$ region at $\alpha=1$ and inside the $68\%$ region with free $\alpha$, and the point for $\wrh=1/3$ lies inside the $68\%$ region in both cases. P-ACT+DESI does not agree. Both the $\wrh=0$ and $\wrh=1/3$ points lie outside its $95\%$ region, the latter only barely with free $\alpha$. Only for $\wrh=1$, which gives $\ns=0.9679$ ($0.9688$), does the gravitational point enter the P-ACT+DESI region, between its $68\%$ and $95\%$ contours. The gravitational benchmark therefore leads to the same conclusion as the rest of the paper, with the same caveat: the only equation of state that works for P-ACT+DESI is one that perturbative oscillations do not produce, and it sits in the region disfavored by the gravitational-wave bound. A similar link between decay rates and $\Nstar$ was explored for no-scale Starobinsky-like models in~\cite{Ellis2015}.

\subsection{Limitations and outlook}
\label{sec:micro_limits}

The reconstruction can be sharpened in three directions. The first concerns \cref{eq:TGamma}, which assumes $H\simeq\Gamma_\phi$ and prompt thermalization. It gives a consistent interpretation of the matter-like marker in \cref{fig:couplings_alpha1,fig:couplings_alphafree}, but for other values of $\wrh$ our effective parametrization does not specify how radiation domination begins. The other two markers therefore show how the predictions change when the same reheating temperature is combined with different effective expansion histories, not self-consistent decay scenarios. Establishing such a connection would require following radiation production and thermalization explicitly.

The second is the microphysics behind the couplings. The conventions we have fixed matter less than the width of the $\Trh$ posterior, which spans $3$--$5$~dex: using $\grh=228.75$ instead of $106.75$ lowers $\Trh$ at fixed $\Gamma_\phi$ by $0.08$~dex, and other normalizations of $g$ and $y$ change them by about $0.15$~dex. With several decay channels each reconstructed coupling becomes an upper bound, and every additional light, minimally coupled scalar adds the same contribution to the gravitational rate~\cite{GorbunovPanin2012}, so that four of them raise $\Trh^{\rm grav}$ by about $0.3$~dex. For large couplings, preheating, backreaction and delayed thermalization are the natural next ingredients.

The third is the data. The coupling posteriors inherit the $\Trh$ posterior, which a flat prior in $\log_{10}\Trh$ maps onto flat priors in $\log_{10}y$ and $\log_{10}g$, so they improve directly with the precision on $\ns$. Including the gravitational-wave bound in the likelihood would remove the low-$\Trh$ part of the P-ACT+DESI posterior and move its couplings upward. Since $\sigma(\ln R_{\rm rad})\simeq\frac12\Nstar^2\sigma(\ns)$, a future measurement with $\sigma(\ns)\simeq10^{-3}$ would determine $\ln R_{\rm rad}$, and hence $\Nstar$, to about $\pm1.6$ $e$-folds. At $\alpha=1$ this would distinguish gravitational decay followed by matter-like reheating ($\ns=0.9623$) from the radiation-like case ($\ns=0.9653$), which differ by three times that uncertainty.

\section{Conclusions}
\label{sec:conclusions}

We have constrained the $\alpha$-Starobinsky model with the reheating history built into the pivot-scale matching. For every sampled point, $\Nstar$ is derived from $(\wrh,\Trh)$ on the exact inflationary background, and the spectra are computed numerically with \texttt{CLASS}, without using slow-roll relations in the inference. We confirm that the CMB is sensitive to reheating only through a single quantity, $\ln R_{\rm rad}\simeq\Nstar-N_{\rm inst}$, and quantify this with the full numerical matching: its precision is set by the constraint on $\ns$, $\sigma(\ln R_{\rm rad})\simeq\frac12\Nstar^2\sigma(\ns)$, which gives $2.5$--$4$ $e$-folds for current data. At $\alpha=1$, Planck is consistent with instantaneous reheating, giving $\ns=0.9656^{+0.0015}_{-0.0017}$ and $\ln R_{\rm rad}=0.4\pm2.7$. In contrast, P-ACT+DESI gives $\ns=0.9689^{+0.0012}_{-0.0016}$ and $\ln R_{\rm rad}=6.5^{+2.5}_{-3.1}$, preferring a stiff phase ($\wrh>0.47$ at 95\%). When $\alpha$ is allowed to vary, the preferred value of $\ns$ increases to $\ns=0.9706^{+0.0016}_{-0.0020}$, with $\log_{10}\alpha=1.0^{+0.5}_{-0.8}$, although the upper range of $\alpha$ remains sensitive to the exclusion of $B$-mode data.

The reheating history also imposes an upper limit on $\ns$. Without a stiff phase, $\ns$ cannot exceed its instantaneous-reheating value, $\ns=0.9653$ at $\alpha=1$ and at most $0.9685$ for $\alpha\le50$, leaving the model outside the $1\sigma$ region of P-ACT-LB. Even for $\wrh=1$, the maximum is $\ns=0.9718$ at $\alpha=1$. If the stiff branch is forbidden, P-ACT+DESI gives $\ns=0.9669^{+0.0008}_{-0.0014}$, and $\Nstar$ stays just below its largest allowed value, $N_{\rm inst}$. This preference is in tension with our approximate gravitational-wave constraint, which excludes roughly half of the preferred stiff region, and conventional oscillatory reheating in the $\alpha$-Starobinsky potential does not provide a sustained stiff expansion history. We therefore interpret the preference for $\wrh>1/3$ as a constraint on the effective post-inflationary expansion history, rather than as evidence for a particular physical realization of stiff reheating. A more complete assessment would require incorporating the gravitational-wave bound directly into the likelihood and including $B$-mode data when $\alpha$ is free, from BICEP/Keck~\cite{BK18} and, in the future, from the Simons Observatory and LiteBIRD~\cite{SimonsObservatory2019,LiteBIRD2023}.

Comparing these results with an analysis in which $\Nstar$ is sampled freely shows that the reheating sector affects the inflationary constraints primarily through the prior it induces on $\Nstar$. At $\alpha=1$, this prior peaks near $N_{\rm inst}$ and narrows the posteriors of $\Nstar$, $\log_{10}V_0$ and $r$ by up to a third, rather than providing additional observational information. When $\alpha$ is free, the physical requirement $\wrh\le1$ excludes the largest values of $\Nstar$ and, for P-ACT+DESI, approximately doubles the preferred $r$. The comparison also has a practical lesson. Since the CMB constrains only $\ln R_{\rm rad}$, the reheating posterior is a thin band in the $(\wrh,\Trh)$ plane that direct sampling explores poorly, and reweighting a chain with free $\Nstar$ is more reliable. For Planck the direct runs underestimate the width of $\ns$ by about a third, but for P-ACT+DESI by less than $10\%$, so our main conclusions are not affected. Finally, we have explored an illustrative connection between the inferred reheating temperatures and perturbative inflaton decay rates. Within the adopted benchmark assumptions, the inflaton mass is $m_\phi\simeq3\times10^{13}$~GeV, with a gravitational-decay reheating temperature of approximately $2\times10^9$~GeV. Planck is compatible with this matter-like reheating benchmark, whereas for P-ACT+DESI the corresponding temperature lies in the preferred region only when combined with a stiff effective expansion history, which perturbative decay in the quadratic minimum does not produce.

\begin{acknowledgments}
KC acknowledges the postdoctoral fellowship from the Secretar\'ia de Ciencias, Humanidades, Tecnolog\'ia e Innovaci\'on (SECIHTI). LEP acknowledges support from JSPS KAKENHI Grant Number JP25KF0278. LEP and JCH would like to thank  UNAM-PAPIIT grant IN114626, ``Estudios de ANPs y Contribuciones a la colaboraci\'on LSST-M\'exico''. GG and JCH acknowledge support from DGAPA-PAPIIT-UNAM grant No.~IN110325, \textit{Estudios en cosmolog\'ia inflacionaria, agujeros negros primordiales y energ\'ia oscura}. The authors gratefully acknowledge the computing facilities provided by the Atocatl cluster at LAMOD-UNAM. LAMOD (\url{http://www.lamod.unam.mx/}) is a collaborative effort between the IA, ICN, and IQ institutes at UNAM. We used ChatGPT (OpenAI) and Claude (Anthropic) to assist with coding, cross-checking calculations, and language editing. The authors reviewed and verified all AI-assisted material and take full responsibility for the content of the paper.
\end{acknowledgments}

\section*{Data availability}
The data underlying this article will be shared on reasonable request to the corresponding author.

\appendix
\crefalias{section}{appendix}
\section{Sampler settings and diagnostics}
\label{app:sampler-validation}

The runs of the reheating model use \texttt{pocoMC} with $n_{\rm active}=256$ particles. Each run stops when the inverse temperature reaches $\beta=1$ and the total effective sample size estimated by the sampler reaches $4096$. The Kish effective sample size of the final weighted sample ranges from $4061$ to $4314$ (\cref{tab:sampler}).

As a test of the direct runs, we repeated the Planck run at $\alpha=1$ with twice as many particles and a different random seed.\footnote{An independent Metropolis--Hastings run of the reheating model, for Planck at $\alpha=1$, was not practical, with an acceptance rate of about $1\%$.} The inflationary parameters changed by less than $0.3\sigma$. Only the prior-dominated posterior of $\Trh$ changed noticeably, with its maximum moving by about one unit in $\log_{10}\Trh$. A run with four times as many particles fills in the low-$\Nstar$ end of the posterior further, but the width of $\ns$ remains below the reweighted value.

\begin{table}[tbp]
\centering
\footnotesize
\renewcommand{\arraystretch}{1.3}\setlength{\tabcolsep}{4pt}
\begin{tabular}{lccccccc}
\toprule
 & \multicolumn{3}{c}{$\alpha=1$} & \multicolumn{3}{c}{$\alpha$ free} & restr. \\
\cmidrule(lr){2-4}\cmidrule(lr){5-7}\cmidrule(lr){8-8}
Combination & iter. & calls & ESS & iter. & calls & ESS & ESS \\
\midrule
Planck & 57 & 602 & 4071 & 57 & 851 & 4221 & 4265 \\
ACT & 45 & 88 & 4070 & 46 & 122 & 4288 & 4223 \\
P-ACT & 48 & 120 & 4125 & 49 & 295 & 4294 & 4256 \\
Planck+DESI & 58 & 520 & 4215 & 58 & 886 & 4149 & 4255 \\
ACT+DESI & 46 & 68 & 4061 & 47 & 115 & 4221 & 4265 \\
P-ACT+DESI & 49 & 101 & 4214 & 50 & 305 & 4314 & 4090 \\
\bottomrule
\end{tabular}
\caption{Sampler diagnostics: number of tempering iterations, likelihood calls (in thousands) and final Kish effective sample size of the weighted posterior sample for each run. The last column gives the effective sample size of the free-$\alpha$ runs with the restrictive prior $\wrh\le1/3$.}
\label{tab:sampler}
\end{table}

\bibliography{alpha_reheating_refs_revised}

@article{Guth1981,
  author  = {{Guth}, Alan H.},
  title   = {{Inflationary universe: A possible solution to the horizon and flatness problems}},
  journal = {\prd},
  year    = 1981,
  volume  = {23},
  number  = {2},
  pages   = {347--356},
  doi     = {10.1103/PhysRevD.23.347}
}

@article{Linde1982,
  author  = {{Linde}, Andrei D.},
  title   = {{A new inflationary universe scenario: A possible solution of the horizon, flatness, homogeneity, isotropy and primordial monopole problems}},
  journal = {Phys. Lett. B},
  year    = 1982,
  volume  = {108},
  number  = {6},
  pages   = {389--393},
  doi     = {10.1016/0370-2693(82)91219-9}
}

@article{AlbrechtSteinhardt1982,
  author  = {{Albrecht}, Andreas and {Steinhardt}, Paul J.},
  title   = {{Cosmology for Grand Unified Theories with Radiatively Induced Symmetry Breaking}},
  journal = {\prl},
  year    = 1982,
  volume  = {48},
  number  = {17},
  pages   = {1220--1223},
  doi     = {10.1103/PhysRevLett.48.1220}
}

@article{MukhanovChibisov1981,
  author  = {{Mukhanov}, V. F. and {Chibisov}, G. V.},
  title   = {{Quantum fluctuations and a nonsingular universe}},
  journal = {JETP Lett.},
  year    = 1981,
  volume  = {33},
  pages   = {532--535}
}

@article{Starobinsky1980,
  author  = {{Starobinsky}, A. A.},
  title   = {{A new type of isotropic cosmological models without singularity}},
  journal = {Phys. Lett. B},
  year    = 1980,
  volume  = {91},
  number  = {1},
  pages   = {99--102},
  doi     = {10.1016/0370-2693(80)90670-X}
}

@article{EncyclopaediaInflationaris,
  author        = {{Martin}, J{\'e}r{\^o}me and {Ringeval}, Christophe and {Vennin}, Vincent},
  title         = {{Encyclop{\ae}dia Inflationaris}},
  journal       = {Phys. Dark Univ.},
  year          = 2014,
  volume        = {5-6},
  pages         = {75--235},
  doi           = {10.1016/j.dark.2014.01.003},
  archivePrefix = {arXiv},
  eprint        = {1303.3787},
  primaryClass  = {astro-ph.CO}
}

@article{PlanckInflation2020,
  author        = {{Akrami}, Y. and others},
  collaboration = {Planck},
  title         = {{Planck 2018 results. X. Constraints on inflation}},
  journal       = {\aap},
  year          = 2020,
  volume        = {641},
  pages         = {A10},
  doi           = {10.1051/0004-6361/201833887},
  archivePrefix = {arXiv},
  eprint        = {1807.06211},
  primaryClass  = {astro-ph.CO}
}

@article{ACTDR6Qu,
  author        = {{Qu}, Frank J. and others},
  collaboration = {ACT},
  title         = {{The Atacama Cosmology Telescope: A Measurement of the DR6 CMB Lensing Power Spectrum and Its Implications for Structure Growth}},
  journal       = {\apj},
  year          = 2024,
  volume        = {962},
  number        = {2},
  pages         = {112},
  doi           = {10.3847/1538-4357/acfe06},
  archivePrefix = {arXiv},
  eprint        = {2304.05202},
  primaryClass  = {astro-ph.CO}
}

@article{ACTDR6Madhavacheril,
  author        = {{Madhavacheril}, Mathew S. and others},
  collaboration = {ACT},
  title         = {{The Atacama Cosmology Telescope: DR6 Gravitational Lensing Map and Cosmological Parameters}},
  journal       = {\apj},
  year          = 2024,
  volume        = {962},
  number        = {2},
  pages         = {113},
  doi           = {10.3847/1538-4357/acff5f},
  archivePrefix = {arXiv},
  eprint        = {2304.05203},
  primaryClass  = {astro-ph.CO}
}

@article{Louis2025,
  author        = {{Louis}, Thibaut and others},
  collaboration = {ACT},
  title         = {{The Atacama Cosmology Telescope: DR6 Power Spectra, Likelihoods and $\Lambda$CDM Parameters}},
  journal       = {\jcap},
  year          = 2025,
  volume        = {11},
  pages         = {062},
  doi           = {10.1088/1475-7516/2025/11/062},
  archivePrefix = {arXiv},
  eprint        = {2503.14452},
  primaryClass  = {astro-ph.CO}
}

@article{DESIDR2,
  author        = {{Abdul Karim}, M. and others},
  collaboration = {DESI},
  title         = {{DESI DR2 results. II. Measurements of baryon acoustic oscillations and cosmological constraints}},
  journal       = {\prd},
  year          = 2025,
  volume        = {112},
  number        = {8},
  pages         = {083515},
  doi           = {10.1103/PhysRevD.112.083515},
  archivePrefix = {arXiv},
  eprint        = {2503.14738},
  primaryClass  = {astro-ph.CO}
}

@article{GarciaQuintero2025,
  author        = {{Garcia-Quintero}, C. and {Noriega}, H.~E. and {de Mattia}, A. and others},
  title         = {{Cosmological Implications of DESI DR2 BAO Measurements in Light of the Latest ACT DR6 CMB Data}},
  journal       = {\prd},
  volume        = {112},
  pages         = {083529},
  year          = 2025,
  eprint        = {2504.18464},
  archivePrefix = {arXiv}
}

@article{Carrion2026,
  author        = {{Carrion}, Karim and {Linares Cede{\~n}o}, Francisco X. and {Germ{\'a}n}, Gabriel and {Hidalgo}, Juan Carlos},
  title         = {{Bayesian analysis of $\alpha$-Starobinsky model with Planck, ACT and DESI data}},
  journal       = {\jcap},
  year          = 2026,
  volume        = {2026},
  number        = {07},
  pages         = {095},
  doi           = {10.1088/1475-7516/2026/07/095},
  archivePrefix = {arXiv},
  eprint        = {2603.25721},
  primaryClass  = {astro-ph.CO}
}

@article{DaiKamionkowskiWang2014,
  author        = {{Dai}, Liang and {Kamionkowski}, Marc and {Wang}, Junpu},
  title         = {{Reheating constraints to inflationary models}},
  journal       = {\prl},
  year          = 2014,
  volume        = {113},
  number        = {4},
  pages         = {041302},
  doi           = {10.1103/PhysRevLett.113.041302},
  archivePrefix = {arXiv},
  eprint        = {1404.6704},
  primaryClass  = {astro-ph.CO}
}

@misc{Lozanov2019,
  author        = {{Lozanov}, Kaloian D.},
  title         = {{Lectures on Reheating after Inflation}},
  year          = 2019,
  eprint        = {1907.04402},
  archivePrefix = {arXiv},
  primaryClass  = {astro-ph.CO}
}

@article{MartinRingevalVennin2024,
  author        = {{Martin}, J{\'e}r{\^o}me and {Ringeval}, Christophe and {Vennin}, Vincent},
  title         = {{Cosmic Inflation at the crossroads}},
  journal       = {\jcap},
  year          = 2024,
  volume        = {2024},
  number        = {07},
  pages         = {087},
  doi           = {10.1088/1475-7516/2024/07/087},
  archivePrefix = {arXiv},
  eprint        = {2404.10647},
  primaryClass  = {astro-ph.CO}
}

@article{Ballardini2025,
  author        = {{Ballardini}, Mario},
  title         = {{Chasing cosmic inflation: constraints for inflationary models and reheating insights}},
  journal       = {\jcap},
  year          = 2025,
  volume        = {2025},
  number        = {01},
  pages         = {116},
  doi           = {10.1088/1475-7516/2025/01/116},
  archivePrefix = {arXiv},
  eprint        = {2408.03321},
  primaryClass  = {astro-ph.CO}
}

@article{Mondal2026,
  author        = {{Mondal}, Rajesh and {Mondal}, Sourav and {Chakraborty}, Ayan},
  title         = {{Constraining reheating temperature, Inflaton-SM coupling and dark matter mass in light of ACT DR6 observations}},
  journal       = {\jcap},
  year          = 2026,
  volume        = {2026},
  number        = {05},
  pages         = {097},
  doi           = {10.1088/1475-7516/2026/05/097},
  archivePrefix = {arXiv},
  eprint        = {2505.13387},
  primaryClass  = {hep-ph}
}

@article{Yi2006,
  author  = {{Yi}, Piljin},
  title   = {{Brane Inflation, Decay of D-Branes, and Reheating of the Universe}},
  journal = {Prog. Theor. Phys. Suppl.},
  year    = 2006,
  volume  = {163},
  pages   = {339--354},
  doi     = {10.1143/PTPS.163.339}
}

@article{BallesterosTamarit2016,
  author        = {{Ballesteros}, Guillermo and {Tamarit}, Carlos},
  title         = {{Radiative plateau inflation}},
  journal       = {JHEP},
  year          = 2016,
  volume        = {2016},
  number        = {02},
  pages         = {153},
  doi           = {10.1007/JHEP02(2016)153},
  archivePrefix = {arXiv},
  eprint        = {1510.05669},
  primaryClass  = {hep-ph}
}

@article{Garcia2020,
  author        = {{Garcia}, Marcos A.~G. and {Kaneta}, Kunio and {Mambrini}, Yann and {Olive}, Keith A.},
  title         = {{Reheating and post-inflationary production of dark matter}},
  journal       = {\prd},
  year          = 2020,
  volume        = {101},
  number        = {12},
  pages         = {123507},
  doi           = {10.1103/PhysRevD.101.123507},
  archivePrefix = {arXiv},
  eprint        = {2004.08404},
  primaryClass  = {hep-ph}
}

@article{Ellis2015,
  author        = {{Ellis}, John and {Garcia}, Marcos A.~G. and {Nanopoulos}, Dimitri V. and {Olive}, Keith A.},
  title         = {{Calculations of Inflaton Decays and Reheating: with Applications to No-Scale Inflation Models}},
  journal       = {\jcap},
  year          = 2015,
  volume        = {07},
  pages         = {050},
  doi           = {10.1088/1475-7516/2015/07/050},
  archivePrefix = {arXiv},
  eprint        = {1505.06986},
  primaryClass  = {hep-ph}
}

@article{GermanPerturbative2025,
  author        = {{Germ{\'a}n}, Gabriel},
  title         = {{Perturbative Analysis of the Reheating Dynamics of $\alpha$-Attractors}},
  journal       = {Int. J. Mod. Phys. D},
  year          = 2025,
  volume        = {34},
  number        = {14},
  pages         = {2550065},
  doi           = {10.1142/S0218271825500658},
  archivePrefix = {arXiv},
  eprint        = {2504.14082},
  primaryClass  = {astro-ph.CO}
}

@article{GermanHidalgo2025,
  author        = {{Germ{\'a}n}, Gabriel and {Hidalgo}, Juan Carlos},
  title         = {{Viability of Generalized $\alpha$-Inflation from Planck, ACT and DESI Data}},
  journal       = {Int. J. Mod. Phys. D},
  volume        = {35},
  number        = {03},
  pages         = {2550098},
  year          = 2026,
  doi           = {10.1142/S0218271825500981},
  eprint        = {2508.01017},
  archivePrefix = {arXiv}
}

@article{KalloshLinde2013,
  author        = {{Kallosh}, Renata and {Linde}, Andrei},
  title         = {{Universality Class in Conformal Inflation}},
  journal       = {\jcap},
  year          = 2013,
  volume        = {07},
  pages         = {002},
  doi           = {10.1088/1475-7516/2013/07/002},
  archivePrefix = {arXiv},
  eprint        = {1306.5220},
  primaryClass  = {hep-th}
}

@article{KalloshLindeRoest2013,
  author        = {{Kallosh}, Renata and {Linde}, Andrei and {Roest}, Diederik},
  title         = {{Superconformal Inflationary $\alpha$-Attractors}},
  journal       = {JHEP},
  year          = 2013,
  volume        = {11},
  pages         = {198},
  doi           = {10.1007/JHEP11(2013)198},
  archivePrefix = {arXiv},
  eprint        = {1311.0472},
  primaryClass  = {hep-th}
}

@article{KalloshLindeRoest2014,
  author        = {{Kallosh}, Renata and {Linde}, Andrei and {Roest}, Diederik},
  title         = {{Large Field Inflation and Double $\alpha$-Attractors}},
  journal       = {JHEP},
  year          = 2014,
  volume        = {08},
  pages         = {052},
  doi           = {10.1007/JHEP08(2014)052},
  archivePrefix = {arXiv},
  eprint        = {1405.3646},
  primaryClass  = {hep-th}
}

@article{CLASS2011,
  author        = {{Blas}, Diego and {Lesgourgues}, Julien and {Tram}, Thomas},
  title         = {{The Cosmic Linear Anisotropy Solving System (CLASS). Part II: Approximation Schemes}},
  journal       = {\jcap},
  year          = 2011,
  volume        = {07},
  pages         = {034},
  doi           = {10.1088/1475-7516/2011/07/034},
  archivePrefix = {arXiv},
  eprint        = {1104.2933},
  primaryClass  = {astro-ph.CO}
}

@article{Cobaya2021,
  author        = {{Torrado}, Jes{\'u}s and {Lewis}, Antony},
  title         = {{Cobaya: Code for Bayesian Analysis of Hierarchical Physical Models}},
  journal       = {\jcap},
  year          = 2021,
  volume        = {05},
  pages         = {057},
  doi           = {10.1088/1475-7516/2021/05/057},
  archivePrefix = {arXiv},
  eprint        = {2005.05290},
  primaryClass  = {astro-ph.IM}
}

@article{LewisGetDist2025,
  author        = {{Lewis}, Antony},
  title         = {{GetDist: a Python Package for Analysing Monte Carlo Samples}},
  journal       = {\jcap},
  year          = 2025,
  volume        = {08},
  pages         = {025},
  doi           = {10.1088/1475-7516/2025/08/025},
  archivePrefix = {arXiv},
  eprint        = {1910.13970},
  primaryClass  = {astro-ph.IM}
}

@article{Karamanis2022pocoMC,
  author        = {{Karamanis}, Minas and {Nabergoj}, David and {Beutler}, Florian and {Peacock}, John A. and {Seljak}, Uro{\v{s}}},
  title         = {{pocoMC}: A Python package for accelerated {Bayesian} inference in astronomy and cosmology},
  journal       = {Journal of Open Source Software},
  year          = 2022,
  volume        = {7},
  number        = {79},
  pages         = {4634},
  doi           = {10.21105/joss.04634},
  archivePrefix = {arXiv},
  eprint        = {2207.05660},
  primaryClass  = {astro-ph.IM}
}

@article{Cook2015,
  author        = {{Cook}, Jessica L. and {Dimastrogiovanni}, Emanuela and {Easson}, Damien A. and {Krauss}, Lawrence M.},
  title         = {{Reheating predictions in single field inflation}},
  journal       = {\jcap},
  year          = 2015,
  volume        = {04},
  pages         = {047},
  doi           = {10.1088/1475-7516/2015/04/047},
  archivePrefix = {arXiv},
  eprint        = {1502.04673},
  primaryClass  = {astro-ph.CO}
}

@article{Ellis2025,
  author        = {{Ellis}, John and {Garcia}, Marcos A.~G. and {Olive}, Keith A. and {Verner}, Sarunas},
  title         = {{Constraints on Attractor Models of Inflation and Reheating from Planck, BICEP/Keck, ACT DR6, and SPT-3G Data}},
  journal       = {\prd},
  volume        = {113},
  pages         = {063571},
  year          = 2026,
  doi           = {10.1103/d35r-7bn8},
  eprint        = {2510.18656},
  archivePrefix = {arXiv}
}

@article{MunozKamionkowski2015,
  author        = {{Mu{\~n}oz}, Julian B. and {Kamionkowski}, Marc},
  title         = {{Equation-of-state parameter for reheating}},
  journal       = {\prd},
  year          = 2015,
  volume        = {91},
  number        = {4},
  pages         = {043521},
  doi           = {10.1103/PhysRevD.91.043521},
  archivePrefix = {arXiv},
  eprint        = {1412.0656},
  primaryClass  = {astro-ph.CO}
}

@article{Turner1983,
  author        = {{Turner}, Michael S.},
  title         = {{Coherent scalar-field oscillations in an expanding universe}},
  journal       = {\prd},
  year          = 1983,
  volume        = {28},
  number        = {6},
  pages         = {1243--1247},
  doi           = {10.1103/PhysRevD.28.1243}
}

@article{ReheatingDynamics2025,
  author        = {{Germ{\'a}n}, Gabriel},
  title         = {{Reheating Dynamics in Inflationary Cosmology: Insights from $\alpha$-Attractor and $\alpha$-Starobinsky Models}},
  journal       = {Eur. Phys. J. Plus},
  volume        = {140},
  pages         = {129},
  year          = 2025,
  eprint        = {2411.01716},
  archivePrefix = {arXiv}
}

@article{ACTConstraints2025,
  author        = {{Yuennan}, Jureeporn and {Atamurotov}, Farruh and {Channuie}, Phongpichit},
  title         = {{ACT Constraints on Marginally Deformed Starobinsky Inflation}},
  journal       = {Phys. Lett. B},
  volume        = {872},
  pages         = {140065},
  year          = 2026,
  eprint        = {2509.23329},
  archivePrefix = {arXiv}
}

@article{StarobinskyTension2026,
  author        = {{Bezerra-Sobrinho}, J. and {Medeiros}, L. G.},
  title         = {{Starobinsky Inflation and the Latest CMB Data: A Subtle Tension?}},
  journal       = {Eur. Phys. J. C},
  volume        = {86},
  pages         = {416},
  year          = 2026,
  eprint        = {2511.06640},
  archivePrefix = {arXiv}
}

@misc{MinimalExtensions2026,
  author        = {{Shobcha}, Nehla and {Risdianto}, Norma Sidik and {Budhi}, Romy Hanang Setya},
  title         = {{Minimal Extensions of the $\alpha$-Starobinsky Model: Reconciling ACT DR6 and Reheating Constraints}},
  year          = 2026,
  eprint        = {2606.24131},
  archivePrefix = {arXiv}
}

@article{QuantumCorrected2025,
  author        = {{Yuennan}, Jureeporn and {Koad}, Patinya and {Atamurotov}, Farruh and {Channuie}, Phongpichit},
  title         = {{Quantum-Corrected $\phi^4$ Inflation in Light of ACT Observations}},
  journal       = {Eur. Phys. J. C},
  volume        = {85},
  pages         = {1307},
  year          = 2025,
  eprint        = {2508.17263},
  archivePrefix = {arXiv}
}

@article{Husdal2016,
  author        = {{Husdal}, Lars},
  title         = {{On Effective Degrees of Freedom in the Early Universe}},
  journal       = {Galaxies},
  volume        = {4},
  pages         = {78},
  year          = 2016,
  eprint        = {1609.04979},
  archivePrefix = {arXiv}
}

@article{Saikawa2018,
  author        = {{Saikawa}, Ken'ichi and {Shirai}, Satoshi},
  title         = {{Primordial gravitational waves, precisely: the role of thermodynamics in the Standard Model}},
  journal       = {\jcap},
  volume        = {05},
  pages         = {035},
  year          = 2018,
  eprint        = {1803.01038},
  archivePrefix = {arXiv}
}

@article{Iacconi2025,
  author        = {{Iacconi}, Laura and {Bhattacharya}, Sukannya and {Fasiello}, Matteo and {Wands}, David},
  title         = {{Closing in on $\alpha$-attractors}},
  journal       = {\jcap},
  volume        = {06},
  pages         = {067},
  year          = 2026,
  eprint        = {2511.14673},
  archivePrefix = {arXiv}
}

@article{Planck2018VI,
  author        = {{Planck Collaboration}},
  title         = {{Planck 2018 results. VI. Cosmological parameters}},
  journal       = {\aap},
  volume        = {641},
  pages         = {A6},
  year          = {2020},
  doi           = {10.1051/0004-6361/201833910},
  eprint        = {1807.06209},
  archivePrefix = {arXiv}
}

@article{GorbunovPanin2011,
  author        = {Gorbunov, D. S. and Panin, A. G.},
  title         = {{Scalaron the mighty: producing dark matter and baryon asymmetry at reheating}},
  journal       = {Phys. Lett. B}, volume = {700}, pages = {157}, year = {2011},
  doi           = {10.1016/j.physletb.2011.04.067}, eprint = {1009.2448}, archivePrefix = {arXiv}
}

@article{GorbunovPanin2012,
  author  = {Gorbunov, D. S. and Panin, A. G.},
  title   = {{Free scalar dark matter candidates in $R^2$-inflation: the light, the heavy and the superheavy}},
  journal = {Phys. Lett. B}, volume = {718}, pages = {15}, year = {2012},
  eprint  = {1201.3539}, archivePrefix = {arXiv}
}

@article{Hawking1982,
  author  = {Hawking, S. W.},
  title   = {{The development of irregularities in a single bubble inflationary universe}},
  journal = {Phys. Lett. B}, volume = {115}, pages = {295}, year = 1982
}

@article{Starobinsky1982,
  author  = {Starobinsky, A. A.},
  title   = {{Dynamics of phase transition in the new inflationary universe scenario and generation of perturbations}},
  journal = {Phys. Lett. B}, volume = {117}, pages = {175}, year = 1982
}

@article{GuthPi1982,
  author  = {Guth, Alan H. and Pi, S. Y.},
  title   = {{Fluctuations in the new inflationary universe}},
  journal = {\prl}, volume = {49}, pages = {1110}, year = 1982
}

@article{BardeenSteinhardtTurner1983,
  author  = {Bardeen, James M. and Steinhardt, Paul J. and Turner, Michael S.},
  title   = {{Spontaneous creation of almost scale-free density perturbations in an inflationary universe}},
  journal = {\prd}, volume = {28}, pages = {679}, year = 1983
}

@article{FKLP2013,
  author  = {Ferrara, Sergio and Kallosh, Renata and Linde, Andrei and Porrati, Massimo},
  title   = {{Minimal supergravity models of inflation}},
  journal = {\prd}, volume = {88}, pages = {085038}, year = 2013, eprint = {1307.7696}, archivePrefix = {arXiv}
}

@article{Calabrese2025,
  author  = {Calabrese, Erminia and others},
  title   = {{The Atacama Cosmology Telescope: DR6 Constraints on Extended Cosmological Models}},
  journal = {\jcap}, volume = {11}, pages = {063}, year = 2025,
  eprint  = {2503.14454}, archivePrefix = {arXiv}
}

@article{Camphuis2025,
  author  = {Camphuis, E. and others},
  title   = {{SPT-3G D1: CMB temperature and polarization power spectra and cosmology from 2019 and 2020 observations of the SPT-3G Main field}},
  journal = {\prd}, volume = {113}, pages = {083504}, year = 2026,
  eprint  = {2506.20707}, archivePrefix = {arXiv}
}

@article{KalloshLindeRoest2025,
  author  = {Kallosh, Renata and Linde, Andrei and Roest, Diederik},
  title   = {{Atacama Cosmology Telescope, South Pole Telescope, and Chaotic Inflation}},
  journal = {\prl}, volume = {135}, pages = {161001}, year = 2025,
  eprint  = {2503.21030}, archivePrefix = {arXiv}
}

@article{DreesXu2025,
  author  = {Drees, Manuel and Xu, Yong},
  title   = {{Refined predictions for Starobinsky inflation and post-inflationary constraints in light of ACT}},
  journal = {Phys. Lett. B}, volume = {867}, pages = {139612}, year = 2025,
  eprint  = {2504.20757}, archivePrefix = {arXiv}
}

@article{Zharov2025,
  author  = {Zharov, D. S. and Sobol, O. O. and Vilchinskii, S. I.},
  title   = {{ACT observations, reheating, and Starobinsky and Higgs inflation}},
  journal = {\prd}, volume = {112}, pages = {023544}, year = 2025,
  doi     = {10.1103/km3q-rm34}, eprint = {2505.01129}, archivePrefix = {arXiv}
}

@article{Haque2025,
  author  = {Haque, Md Riajul and Pal, Sourav and Paul, Debarun},
  title   = {{ACT DR6 insights on $\alpha$-attractor inflationary models and reheating}},
  journal = {\prd}, volume = {114}, pages = {023521}, year = 2026,
  doi     = {10.1103/1kj4-s4ts}, eprint = {2505.01517}, archivePrefix = {arXiv}
}

@misc{Dorsch2026,
  author = {Dorsch, Gl{\'a}uber C. and Le{\~a}o, Lucas P. C. and Alves, Lucas M. B. and Miranda, Luiz C. A. and Sena, Beatriz M. D.},
  title  = {{Reheating matters: Starobinsky inflation in light of joint CMB+BAO results and gravitational-wave forecasts}},
  year   = 2026, eprint = {2606.28502}, archivePrefix = {arXiv}
}

@article{Eroncel2025,
  author  = {Er{\"o}ncel, Cem and Gouttenoire, Yann and Sato, Ryosuke and Servant, G{\'e}raldine and Simakachorn, Peera},
  title   = {{Universal Bound on the Duration of a Kination Era}},
  journal = {\prl}, volume = {135}, pages = {101002}, year = 2025,
  doi     = {10.1103/k7ty-gwjg}, eprint = {2501.17226}, archivePrefix = {arXiv}
}

@article{FigueroaTanin2019,
  author  = {Figueroa, Daniel G. and Tanin, Erwin H.},
  title   = {{Ability of LIGO and LISA to probe the equation of state of the early Universe}},
  journal = {\jcap}, volume = {08}, pages = {011}, year = 2019, eprint = {1905.11960}, archivePrefix = {arXiv}
}

@article{Giovannini1998,
  author  = {Giovannini, Massimo},
  title   = {{Gravitational wave constraints on postinflationary phases stiffer than radiation}},
  journal = {\prd}, volume = {58}, pages = {083504}, year = 1998, eprint = {gr-qc/9806329}
}

@article{BoyleSteinhardt2008,
  author  = {Boyle, Latham A. and Steinhardt, Paul J.},
  title   = {{Probing the early universe with inflationary gravitational waves}},
  journal = {\prd}, volume = {77}, pages = {063504}, year = 2008, eprint = {astro-ph/0512014}
}

@article{MartinRingeval2010,
  author  = {Martin, J{\'e}r{\^o}me and Ringeval, Christophe},
  title   = {{First CMB constraints on the inflationary reheating temperature}},
  journal = {\prd}, volume = {82}, pages = {023511}, year = 2010, eprint = {1004.5525}, archivePrefix = {arXiv}
}

@article{LiddleLeach2003,
  author  = {Liddle, Andrew R. and Leach, Samuel M.},
  title   = {{How long before the end of inflation were observable perturbations produced?}},
  journal = {\prd}, volume = {68}, pages = {103503}, year = 2003, eprint = {astro-ph/0305263}
}

@article{DodelsonHui2003,
  author  = {Dodelson, Scott and Hui, Lam},
  title   = {{A horizon ratio bound for inflationary fluctuations}},
  journal = {\prl}, volume = {91}, pages = {131301}, year = 2003, eprint = {astro-ph/0305113}
}

@article{Rosenberg2022,
  author  = {Rosenberg, Erik and Gratton, Steven and Efstathiou, George},
  title   = {{CMB power spectra and cosmological parameters from Planck PR4 with CamSpec}},
  journal = {\mnras}, volume = {517}, pages = {4620}, year = 2022, eprint = {2205.10869}, archivePrefix = {arXiv}
}

@article{Carron2022,
  author  = {Carron, Julien and Mirmelstein, Mark and Lewis, Antony},
  title   = {{CMB lensing from Planck PR4 maps}},
  journal = {\jcap}, volume = {09}, pages = {039}, year = 2022, eprint = {2206.07773}, archivePrefix = {arXiv}
}

@article{Planck2018V,
  author  = {{Planck Collaboration}},
  title   = {{Planck 2018 results. V. CMB power spectra and likelihoods}},
  journal = {\aap}, volume = {641}, pages = {A5}, year = 2020, eprint = {1907.12875}, archivePrefix = {arXiv}
}

@article{Pagano2020,
  author  = {Pagano, L. and others},
  title   = {{Reionization optical depth determination from Planck HFI data with ten percent accuracy}},
  journal = {\aap}, volume = {635}, pages = {A99}, year = 2020, eprint = {1908.09856}, archivePrefix = {arXiv}
}

@misc{Lesgourgues2011,
  author = {Lesgourgues, Julien},
  title  = {{The Cosmic Linear Anisotropy Solving System (CLASS) I: Overview}},
  year   = 2011, eprint = {1104.2932}, archivePrefix = {arXiv}
}

@article{Karamanis2022MNRAS,
  author  = {Karamanis, Minas and Beutler, Florian and Peacock, John A. and Nabergoj, David and Seljak, Uro{\v s}},
  title   = {{Accelerating astronomical and cosmological inference with preconditioned Monte Carlo}},
  journal = {\mnras}, volume = {516}, pages = {1644}, year = 2022, eprint = {2207.05652}, archivePrefix = {arXiv}
}

@misc{KaramanisSeljak2024,
  author = {Karamanis, Minas and Seljak, Uro{\v s}},
  title  = {{Persistent Sampling: Enhancing the Efficiency of Sequential Monte Carlo}},
  year   = 2024, eprint = {2407.20722}, archivePrefix = {arXiv}
}

@article{Hasegawa2019,
  author  = {Hasegawa, Takuya and others},
  title   = {{MeV-scale reheating temperature and thermalization of oscillating neutrinos by radiative and hadronic decays of massive particles}},
  journal = {\jcap}, volume = {12}, pages = {012}, year = 2019, eprint = {1908.10189}, archivePrefix = {arXiv}
}

@article{deSalas2015,
  author  = {de Salas, P. F. and Lattanzi, M. and Mangano, G. and Miele, G. and Pastor, S. and Pisanti, O.},
  title   = {{Bounds on very low reheating scenarios after Planck}},
  journal = {\prd}, volume = {92}, pages = {123534}, year = 2015, eprint = {1511.00672}, archivePrefix = {arXiv}
}

@article{UenoYamamoto2016,
  author  = {Ueno, Yoshiki and Yamamoto, Kazuhiro},
  title   = {{Constraints on $\alpha$-attractor inflation and reheating}},
  journal = {\prd}, volume = {93}, pages = {083524}, year = 2016, eprint = {1602.07427}, archivePrefix = {arXiv}
}

@article{BK18,
  author  = {{BICEP/Keck Collaboration}},
  title   = {{Improved Constraints on Primordial Gravitational Waves using Planck, WMAP, and BICEP/Keck Observations through the 2018 Observing Season}},
  journal = {\prl}, volume = {127}, pages = {151301}, year = 2021, eprint = {2110.00483}, archivePrefix = {arXiv}
}

@article{Hidalgo:2011fj,
    author = "Hidalgo, J. C. and Urena-Lopez, L. Arturo and Liddle, Andrew R.",
    title = "{Unification models with reheating via Primordial Black Holes}",
    eprint = "1107.5669",
    archivePrefix = "arXiv",
    primaryClass = "astro-ph.CO",
    doi = "10.1103/PhysRevD.85.044055",
    journal = "Phys. Rev. D",
    volume = "85",
    pages = "044055",
    year = "2012"
}

@article{Allahverdi2010,
  author        = {{Allahverdi}, Rouzbeh and {Brandenberger}, Robert and {Cyr-Racine}, Francis-Yan and {Mazumdar}, Anupam},
  title         = {{Reheating in Inflationary Cosmology: Theory and Applications}},
  journal       = {Ann. Rev. Nucl. Part. Sci.},
  year          = 2010,
  volume        = {60},
  pages         = {27--51},
  doi           = {10.1146/annurev.nucl.012809.104511},
  eprint        = {1001.2600},
  archivePrefix = {arXiv}
}

@article{SimonsObservatory2019,
  author        = {{Simons Observatory Collaboration}},
  title         = {{The Simons Observatory: Science goals and forecasts}},
  journal       = {\jcap},
  year          = 2019,
  volume        = {02},
  pages         = {056},
  doi           = {10.1088/1475-7516/2019/02/056},
  eprint        = {1808.07445},
  archivePrefix = {arXiv}
}

@article{LiteBIRD2023,
  author        = {{LiteBIRD Collaboration}},
  title         = {{Probing Cosmic Inflation with the LiteBIRD Cosmic Microwave Background Polarization Survey}},
  journal       = {PTEP},
  year          = 2023,
  volume        = {2023},
  number        = {4},
  pages         = {042F01},
  doi           = {10.1093/ptep/ptac150},
  eprint        = {2202.02773},
  archivePrefix = {arXiv}
}

\end{document}